\documentclass[10pt,english,aps,prb,amsmath,amssymb,showkeys,superscriptaddress,twocolumn,showpacs,floatfix,longbibliography]{revtex4-2}
\usepackage{graphicx}
\usepackage{dcolumn}
\usepackage[colorlinks=true,linkcolor=blue ,citecolor=blue,urlcolor=blue]{hyperref}
\usepackage{multirow}
\usepackage[usenames,dvipsnames]{xcolor}
\usepackage{soul}
\usepackage{braket}
\usepackage{bm}
\usepackage{nicefrac}
\usepackage{siunitx}
\usepackage{color,transparent}
\usepackage[utf8]{inputenc}
\usepackage{upgreek}
\usepackage[capitalize]{cleveref}
\usepackage{cancel}

\renewcommand{\deg}{$^{\circ}$}

\newcommand{\VEC}[1]{\bm{#1}}

\definecolor{teal}{rgb}{0,0.502,0.502}

\DeclareSIUnit\ML{ML}
\DeclareSIUnit\MLs{MLs}
\DeclareSIUnit\meVA{meV\angstrom^2}

\newcommand{\CharlesUni}{Charles University, Faculty of Mathematics and Physics, Department of Condensed Matter Physics, Ke Karlovu 5, 121 16, Praha, Czech Republic}
\newcommand{\PSISciComp}{PSI Center for Scientific Computing, Theory, and Data, 5232 Villigen PSI, Switzerland}
\newcommand{\MARVEL}{National Centre for Computational Design and Discovery of Novel Materials (MARVEL), 5232 Villigen PSI, Switzerland}
\newcommand{\CBPF}{Centro Brasileiro de Pesquisas Físicas (CBPF), Rua Dr. Xavier Sigaud 150, Urca, Rio de Janeiro - RJ, 22290-180, Brazil}
\newcommand{\THEOS}{Theory and Simulation of Materials (THEOS), \'Ecole Polytechnique F\'ed\'erale de Lausanne, 1015 Lausanne, Switzerland}
\newcommand{\JCNSMLZ}{J\"{u}lich Centre for Neutron Science (JCNS) at Heinz Maier-Leibnitz Zentrum (MLZ), Forschungszentrum J\"{u}lich GmbH, Lichtenbergstrasse 1, D-85747 Garching, Germany}
\newcommand{\JCNSILL}{J\"{u}lich Centre for Neutron Science (JCNS) at Institut Laue Langeving (ILL), Forschungszentrum J\"{u}lich GmbH, 71 Avenue des Martyrs, F-38000 Grenoble, France}
\newcommand{\UniGrenoble}{Universit\'e Grenoble Alpes, CEA, IRIG, MEM, MDN, F-38000 Grenoble, France}

\begin{document}

\title{Spin structures and phase diagrams of the $\text{spin-}\frac{5}{2}$ triangular-lattice antiferromagnet Na$_2$BaMn(PO$_4$)$_2$ under magnetic field}

\author{N. Biniskos}
\email{nikolaos.biniskos@matfyz.cuni.cz}
\affiliation{\CharlesUni}

\author{F. J. dos Santos}
\email{flaviano@cbpf.br}
\affiliation{\PSISciComp}
\affiliation{\MARVEL}
\affiliation{\CBPF}

\author{M. Stekiel}
\email{m.stekiel@fz-juelich.de}
\affiliation{\JCNSMLZ}

\author{K. Schmalzl}
\affiliation{\JCNSILL}

\author{E. Ressouche}
\affiliation{\UniGrenoble}

\author{D. Svit\'ak}
\affiliation{\CharlesUni}

\author{A. Labh}
\affiliation{\CharlesUni}

\author{M. Vali\v{s}ka}
\affiliation{\CharlesUni}

\author{N. Marzari}
\affiliation{\PSISciComp}
\affiliation{\MARVEL}
\affiliation{\THEOS}

\author{P. \v{C}erm\'ak}
\affiliation{\CharlesUni}

\date{\today}

\begin{abstract}
We combine single-crystal neutron diffraction studies and Monte Carlo simulations to determine the spin structures and finite-temperature phase diagram of the spin-5/2 triangular-lattice antiferromagnet Na$_2$BaMn(PO$_4$)$_2$ in magnetic field.
With the application of a magnetic field in two different directions, namely along the $c$-axis and in the $ab$-plane of the trigonal symmetry, we track the evolution of the spin structure through changes of the magnetic propagation vector.
We account for these results with a minimal Heisenberg Hamiltonian that includes easy-axis anisotropy and weak, frustrated interlayer couplings in addition to intralayer exchange.
Guided by representation analysis, we refine symmetry-allowed modes to the measured intensities and obtain the spin structures for all field-induced phases, which we compare quantitatively with simulated configurations.
Taken together, our measurements and simulations show that frustrated interlayer exchange -- rather than purely two-dimensional (2D) physics -- organizes the unexpectedly rich field-induced phases of Na$_2$BaMn(PO$_4$)$_2$. 
\end{abstract}

\date{\today}

\maketitle

\section{Introduction}
\label{sec:intro}

Frustrated triangular-lattice antiferromagnets (TLAs) are considered as model systems in which exotic states can emerge.
For example, the suppression of exchange interactions due to geometric frustration and a low quantum spin number ($S = 1/2$) can result in strong quantum spin fluctuations at low temperatures that stabilize a quantum spin liquid state~\cite{Savary_2017}.
Such states exhibit several appealing properties in condensed matter physics like quasiparticle fractionalization~\cite{Balents_2010}, long-range entanglement~\cite{Broholm_2020}, and topological order~\cite{Wen_1991}.
Novel phases of matter and interesting physical properties are not restricted only to frustrated magnet systems in the quantum limit (low spin), but can also emerge in the classical limit ($S \geq  1$)~\cite{Collins_1997,Tapp_2017,Lee_2020} and by applying an external magnetic field~\cite{Matsuda_2010,Lee_2014,Rosales_2013,Seabra_2016}.

In this context, the equilateral TLAs Na$_2$BaM(PO$_4$)$_2$ (with $\text{M} = \text{Co, Ni, Mn}$) are good candidate materials to explore phenomena related to quantum magnetism, since the substitution of the transition metal ion leads to drastic changes in the quantum spin number~\cite{Zhong_2019,Li_2019,Kim_2022}.
The ground state of the phosphate with the lowest spin ($S = 1/2$), Na$_2$BaCo(PO$_4$)$_2$, is proposed to form a Y-like spin supersolid phase in zero magnetic field~\cite{Xiang_2024,Gao_2022}, and is considered an ideal material where a spin-1/2 easy-axis XXZ model can be applied~\cite{Sheng_2022}.
In addition, the spin supersolidity is connected to a giant magnetocaloric effect which is observed during the demagnetization cooling process~\cite{Xiang_2024}.
On the contrary, Na$_2$BaNi(PO$_4$)$_2$ where $S = 1$, is assumed to exhibit a spin nematic supersolid ground state that can be modeled by a spin Hamiltonian including a single-ion anisotropic term and nearest-neighbor XXZ-type exchange interactions~\cite{Sheng_2025}. 
Although this scenario is primarily supported by field-induced behavior and magnon-pair condensation at low temperatures~\cite{Sheng_2025}, the precise nature of the zero-field ground state remains debated~\cite{huang_2025u}.
Na$_2$BaMn(PO$_4$)$_2$ is expected to be the classical counterpart of the TLA series due to the high quantum spin number ($S = 5/2$) of the Mn$^{+2}$ ions~\cite{Kim_2022}.
Also, it is worth mentioning that all compounds within the series manifest multiple field-induced phase transitions~\cite{Li_2020,Li_2019,Kim_2022}.
Above a critical field applied along the $c$-axis, a magnetization plateau appears in a finite range of the external magnetic field where the magnetization reaches 1/3 of the saturation magnetization $M_\text{sat}$. Such 1/3--magnetization plateaus were first observed in GdPd$_2$Al$_3$~\cite{Kitazawa_1999} and later in other TLAs such as RbFe(MoO$_4$)$_2$~\cite{Svistov_2006} and Ba$_3$MnNb$_2$O$_9$~\cite{Lee_2014}. 
While for Na$_2$BaMn(PO$_4$)$_2$ the magnetic structure at zero field was recently investigated with neutron powder diffraction measurements~\cite{Zhang_2024}, the spin configurations of the field-induced phases and a minimal model Hamiltonian that describes the magnetic ground state remain unknown.

In this article, we report the synthesis of high quality crystals of Na$_2$BaMn(PO$_4$)$_2$ and unpolarized single-crystal neutron diffraction measurements.
We investigated the field-induced phase transitions by recording magnetic reflections as a function of temperature ($T$) in different magnetic fields ($H$), applied along the $c$-axis and in the $ab$-plane of trigonal symmetry.
By identifying the critical fields obtained from heat capacity and neutron diffraction measurements, we mapped the Na$_2$BaMn(PO$_4$)$_2$ phase diagrams for the two magnetic field directions.
With a combination of spin dynamics and Monte Carlo simulations we attempt to determine the magnetic ground state of the system, reproduce the phase diagrams and track the changes in the spin structure versus the applied magnetic field.
In order to compare qualitatively the orientation of the ordered moments in the various field-induced transitions, we refine the obtained magnetic structures based on the integrated intensities of the magnetic reflections.

\section{Methods}
\label{sec:methods}

\subsection{Experimental details}

Na$_2$BaMn(PO$_4$)$_2$ single crystals were grown by the high temperature flux method~\cite{Zhong_2019,Kim_2022} and were characterized with single crystal X-ray diffraction and specific heat measurements.
Single crystal X-ray diffraction measurements were performed on a Rigaku XtaLAB Synergy-S diffractometer, using a Mo X-ray source providing a monochromatic beam with a wavelength of 0.71\,$\text{\AA}$ and a two-dimensional HyPix-Arc 150$^{\circ}$ detector.
The observed reflections were indexed and integrated using the data reduction program \textsc{CrysalisPro}~\cite{Crysalis}.
The structure solution and refinement was performed with SHELX~\cite{shelx}.
Heat capacity was measured on a Quantum Design Physical Property Measurement System (PPMS) equipped with the Heat Capacity option. 
Data were collected on a single crystal of Na$_2$BaMn(PO$_4$)$_2$ with a mass of about 0.6\,mg, employing both the semi-adiabatic time-relaxation technique and the long-pulse method described in Ref.~\cite{Scheie_2018}.

A parallelepiped Na$_2$BaMn(PO$_4$)$_2$ single crystal with mass 37.7\,mg was mounted on a copper sample holder and was oriented in the $(hh0)$/$(00l)$ scattering plane of the trigonal lattice.
In this article, we use the hexagonal coordinate system and the scattering vector $\VEC Q$ is expressed in $\VEC Q = (Q_{h}, Q_{k}, Q_{l})$ given in reciprocal lattice units (r.l.u.). 
The relation between $\VEC Q$ and the propagation vector $\VEC k$ is given by $\VEC Q = \VEC G + \VEC k$, where $\VEC G$ is at the $\Gamma$-point of the reciprocal lattice.

Single-crystal neutron diffraction measurements were carried out at the Institut Laue-Langevin (ILL), in Grenoble, France.
Data were obtained on the two-axis thermal neutron diffractometer D23~\cite{D23-LSF}.
A copper monochromator provided an unpolarized neutron beam with a wavelength of 1.282\,$\text{\AA}$. 
In order to reach sub-Kelvin temperatures, the sample holder was attached to the mixing chamber of a dilution stick. 
To investigate the evolution of the spin structure under magnetic field, measurements were performed in two geometries: with a 15\,T vertical field magnet, providing magnetic field within the $ab$-plane, and with a 3.8\,T horizontal field magnet, providing field along the $c$-axis.
In both geometries, data at zero field were collected and combined to refine the zero-field magnetic structures at 600 and 1200\,mK.
The various spin structures were refined using Mag2POL software~\cite{Qureshi2019}.

\subsection{Theoretical framework}

The effective interactions between the magnetic moments of the Mn sites of Na$_2$BaMn(PO$_4$)$_2$ were modeled using a classical Heisenberg Hamiltonian:
\begin{equation}
    \mathcal H = - \sum_{ij} J_{ij} \bm{S}_i \cdot \bm{S}_j - k^{c} \sum_{i}  (S^c_i)^2 .
\end{equation}
Here, $J_{ij}$ are the magnetic exchange interaction parameters and $k^c$ corresponds to the single-ion anisotropy along the $c$-axis.
Negative $J$ values characterize AFM coupling.
We will discuss the choice of parameters in detail in Section\,\ref{sec:results}, together with the experimental findings.

For this model, the ground state spin configuration was determined through spin dynamics by solving the Landau–Lifshitz–Gilbert equation using the \texttt{Spirit package}~\cite{muller_spirit:_2019}. 
The spin-wave spectrum was determined using linear spin-wave theory for noncollinear magnets~\cite{dos_santos_spin-resolved_2018}. 
Finally, Monte Carlo simulations were employed to determine the mechanical statistical properties of the system using the \texttt{Vampire package}~\cite{evans_atomistic_2014, vampire}. 
We used a $3\times 3\times 5$ macrocell with periodic boundaries and an equilibration time of 5000 Monte Carlo steps, and statistical averaging also over 5000 steps.

\section{Results and discussion}
\label{sec:results}

\subsection{Crystal structure}

\begin{figure}[ht]
    \centering
    \includegraphics[width=\columnwidth]{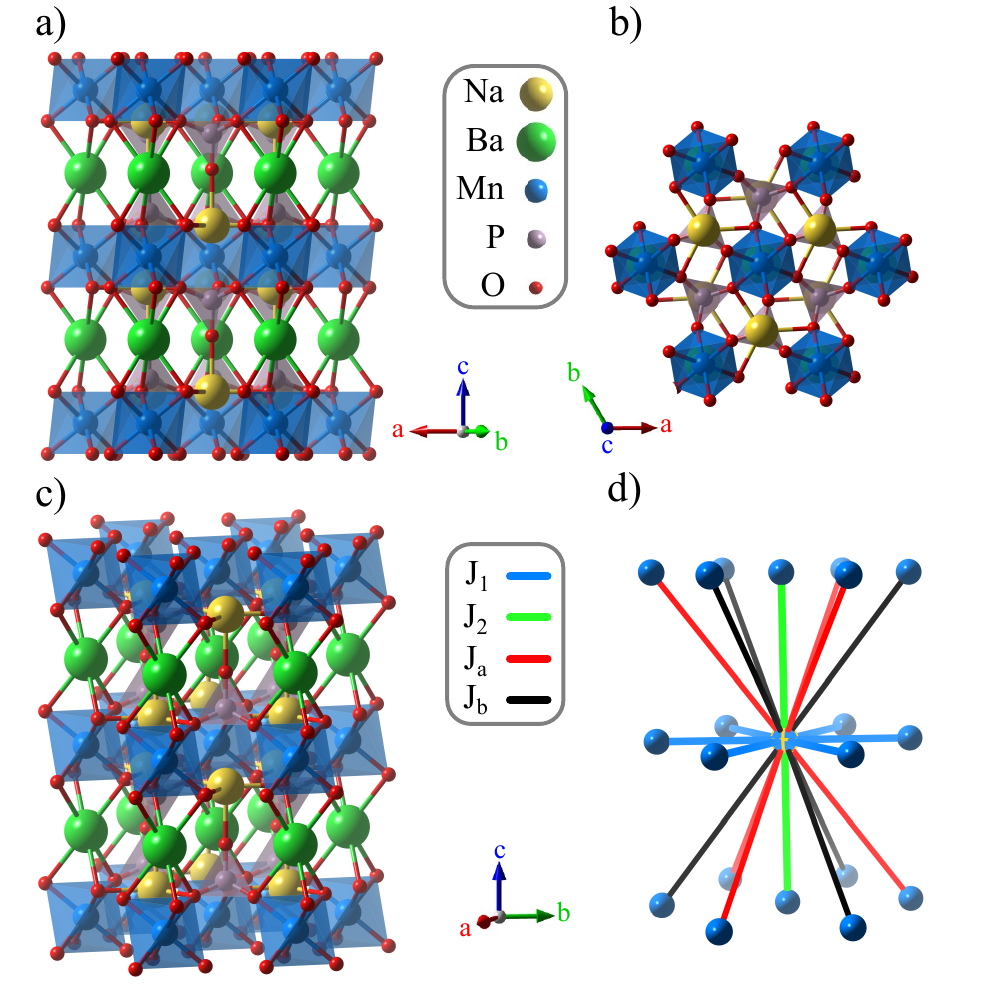}
    \caption{Crystal structure of Na$_2$BaMn(PO$_4$)$_2$ as determined by single crystal X-ray diffraction. 
    View along the (a) in-plane direction, highlighting the layered structure. 
    (b) View along the $c$-axis. 
    (c,d) General view highlighting the connectivity of Mn ions, where (d) shows the Mn--Mn bonds used in theoretical modeling of the spin interactions, in-plane J$_1$ (bond length is $5.37\,$\AA), out-of-plane J$_2$ ($7.094\,$\AA), and diagonal out-of-plane J$_a$ and J$_b$ (both $8.9\,$\AA). Even though the bond length for J$_a$ and J$_b$ is the same, the bonds are not equivalent in the $P\bar{3}$ space group.}
    \label{fig:crystal-structure}
\end{figure}

\begin{table*}[thb]
\begin{center}
\begin{ruledtabular}
\caption{Crystal structure parameters of Na$_2$BaMn(PO$_4$)$_2$ based on the refinement of single crystal X-ray diffraction data at 296\,K. 
The space group is $P\bar{3}$ and the lattice parameters are $a=5.37304(4)$\,$\text{\AA}$, $c=7.0944(10)$\,$\text{\AA}$. 
The number of measured, independent, and observed reflections are 304432, 2564, and 2444, respectively. 
The reflection merge factor is $R_\mathrm{int}$=6.45 and observed reflections are with $I>2\sigma(I)$. 
The refinement was performed on $F^2$ with $R_\mathrm{obs}$=1.63 and $wR_\mathrm{obs}$=4.92. The columns contain the name of the elements, Wyckoff site symbols, positions in crystal coordinates, and displacement parameters, respectively. 
Numbers without errors are restricted by site symmetry.}
\label{tab:crystal-structure}
\begin{tabular}{c c c c c c c c c c}

    Atom & Site & & Position & & & U (\AA$^2$) & & & \\
     &  & x & y & z & U$_{11}$=U$_{22}$ & U$_{33}$ & U$_{12}$ & U$_{23}$=-U$_{13}$ & U$_\mathrm{eq}$
\\ \hline
    Mn & 1$a$ & 
    0 & 0 & 0 &   
    0.00758(3) &   
    0.00663(4) &   
    0.00379(1) &   
    0 &   
    0.00726(2)   
\\
    Ba & 1$b$ &
    0 & 0 & $\frac{1}{2}$ &   
    0.01385(2) &   
    0.00636(2) &   
    0.00693(1) &   
    0 &   
    0.01135(2)   
\\
    Na & 2$d$ &
    $\frac{1}{3}$ & $\frac{2}{3}$ & 0.18600(14) &   
    0.01547(13) &   
    0.0217(3) &   
    0.00773(7) &   
    0 &   
    0.01755(10)   
\\
    P & 2$d$ &
    $\frac{2}{3}$ & $\frac{1}{3}$ & 0.26286(4) &   
    0.00649(4) &   
    0.00636(6) &   
    0.00324(2) &   
    0 &   
    0.00645(3)   
\\
    O & 2$d$ &
    $\frac{2}{3}$ & $\frac{1}{3}$ & 0.47708(12) &   
    0.01831(18) &   
    0.00663(17) &   
    0.00915(9) &   
    0 &   
    0.01441(11)   
\\
    O & 6$i$ &
    0.36193(12) & 0.23574(15) & 0.18706(10) &   
    0.00845(12) &   
    0.01453(18) &   
    0.00543(12) &   
    -0.00322(15) &   
    0.01347(7)   
\\ 
\end{tabular}
\end{ruledtabular}
\end{center}
\end{table*}

The crystal structure of the manganese phosphate Na$_2$BaMn(PO$_4$)$_2$ was determined by single crystal X-ray diffraction at room temperature.
The compound crystallizes in the trigonal lattice with space group $P\bar{3}$ (147) and lattice parameters $a=5.37304(4)$\,$\text{\AA}$ and $c=7.0944(10)$\,$\text{\AA}$.
The structure is presented in Figure~\ref{fig:crystal-structure} and details of the refinement are given in Table~\ref{tab:crystal-structure}.
The building blocks of the crystal structure are regular MnO$_6$ octahedra and PO$_4$ tetrahedra with interstitial Na and Ba ions.
MnO$_6$ octahedra are corner connected through PO$_4$ groups and form layers stacked along the [001] direction.
The Na and Ba ions fill the interstitial positions and bond adjacent MnO layers.

All measured crystals exhibit merohedral twinning with apparent $P\bar{3}m1$ symmetry, where the mirror plane $m_{100}$ defines the twinning law.
Regarding the atomic positions, the difference in assigning either space group concerns only the oxygen at the $6i$ site, where constraints of the $P\bar{3}m1$ space group give $y(\mathrm{O_{6i}})=\frac{1}{2}x(\mathrm{O_{6i}})$, while the $P\bar3$ space group sets the $y$ coordinate free.
The refinement with the $P\bar{3}m1$ space group results in five times larger displacement parameters of $\mathrm{O_{6i}}$ and three times higher R factors than the refinement with the $P\bar{3}$ structure.
Releasing the constraints and refining the structure with $P\bar{3}$ space group yields excellent quality factors and gives the twinning ratio of 70:30.
The $y$ coordinate of position $\mathrm{O_{6i}}$ differs significantly from the $y=\frac{1}{2}x$ constraint ($y\approx 0.24$ vs. constrained $y\approx0.18$) and corresponds to the rotation of the MnO$_6$ octahedra by an angle of 9.9\deg\ from their positions in the $P\bar{3}m1$ space group.
The twinning is consistent with the recently reported high temperature transition in Na$_2$BaMn(PO$_4$)$_2$~\cite{Kajita_2024} and explains the inconsistencies between the reports on the $P\bar{3}$~\cite{Daisuke_2014,Zhang_2024,Zhang_2024b,Kajita_2024} and $P\bar{3}m1$~\cite{Nenert2020,Kim_2022} space groups.
It is worth mentioning that a similar twinning scheme was recently considered for Na$_2$BaCo(PO$_4$)$_2$~\cite{Woodland_2025}.


\subsection{Field-induced phase transitions}
\label{sec:phase-transitions}

\begin{figure}[h]
    \includegraphics[width=\columnwidth]{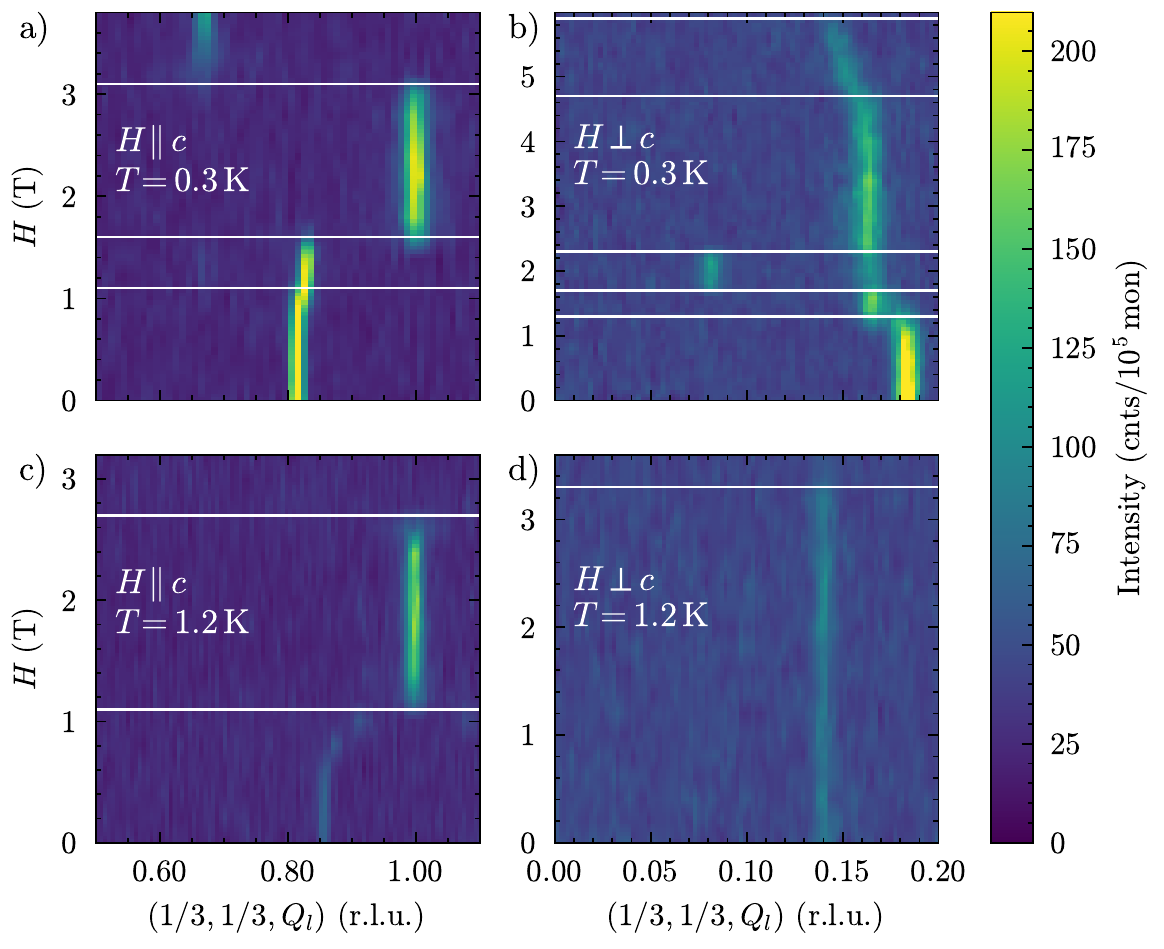}
    \caption
    {Color-coded intensity plots of single crystal neutron diffraction data of Na$_2$BaMn(PO$_4$)$_2$ collected at 300 and 1200\,mK as a function of $\VEC Q = (1/3, 1/3, Q_{l})$ and applied magnetic field along (a), (c) the $c$-axis and (b), (d) in the $ab$-plane. 
    White horizontal lines represent the identified phase boundaries.}
    \label{fig:rawdata2}
\end{figure}

Zero magnetic field heat capacity measurements on single crystals~\cite{Kim_2022} and neutron powder diffraction experiments~\cite{Zhang_2024} revealed that long-ranged antiferromagnetic (AFM) order appears in Na$_2$BaMn(PO$_4$)$_2$ at low temperatures.
The system at zero field undergoes two successive magnetic transitions at $T_{N_2} \approx 1.28$\,K (AFM2) and $T_{N_1} \approx 1.13$\,K (AFM1) into antiferromagnetic phases that are indexed with the modulation vector $\VEC k_\mathrm{AFM2} = (1/3, 1/3, 0.139)$ and $\VEC k_\mathrm{AFM1} = (1/3, 1/3, 0.187)$, respectively~\cite{Zhang_2024}.
Consistently our neutron data in Fig.~\ref{fig:rawdata2} indicate the occurrence of spin ordering at zero field for $T < T_{N_1}$ with $\VEC k_\mathrm{AFM1} = (1/3, 1/3, 0.187)$ and for $T_{N_1} \leq T \leq T_{N_2}$ with $\VEC k_\mathrm{AFM2} = (1/3, 1/3, 0.147)$.
The significant difference in $\VEC k_\mathrm{AFM2}$ between our results and Ref.~\cite{Zhang_2024} is ascribed to the temperature dependence of the propagation vector within this phase.
We also note that due to the blockage of the bulky horizontal magnet several magnetic reflections constrained in the $(hhl)$ plane are not accessible. 
Therefore, data could not be obtained for $-0.5 < Q_l < 0.5$\,r.l.u. for $\VEC H \parallel \VEC{\hat{c}}$ (see Figs~\ref{fig:rawdata2}(a),(c)).

\begin{figure}[ht]
    \includegraphics[width=\columnwidth]{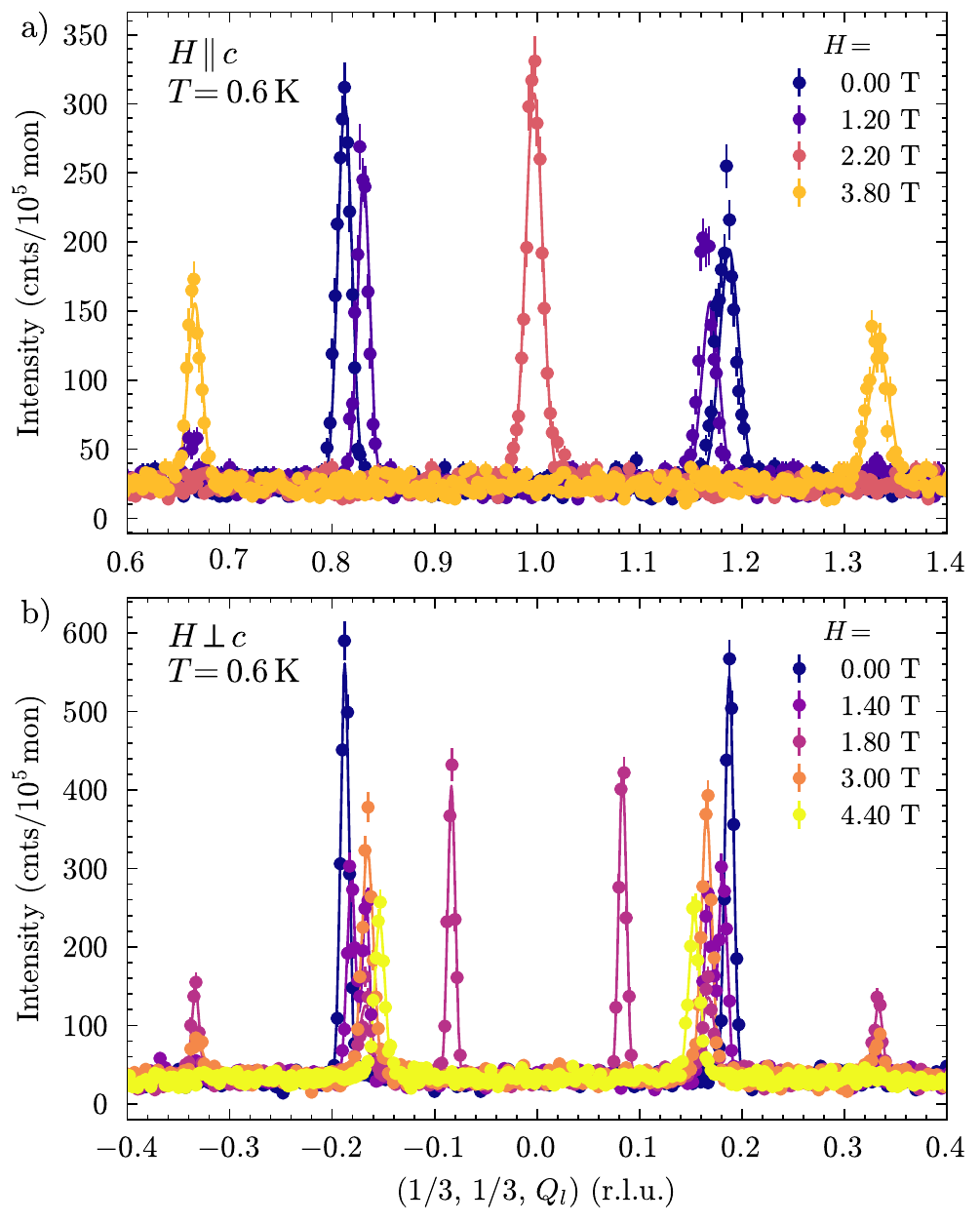}
    \caption
    {Representative reciprocal space scans at $\VEC Q = (1/3, 1/3, Q_{l})$ measured at 600\,mK, under different applied fields applied along (a) the $c$-axis and (b) in the $ab$-plane. Lines represent fits with Gaussian functions on top of a constant background.
    }
    \label{fig:rawdata3}
\end{figure}

Based on $\chi_\text{ac}$ magnetic susceptibility measurements on single-crystal samples~\cite{Kim_2022}, when a magnetic field is applied parallel or perpendicular to the $c$-axis of the trigonal symmetry, multiple field-induced phase transitions occur, whose spin configurations are still unexplored to our knowledge.
Therefore, a first essential step for determining the spin arrangement of the various spin structures is to index their magnetic propagation vector.
To this aim, we performed reciprocal space scans for magnetic fields $\VEC H \parallel \VEC{\hat{c}}$ and $\VEC H \perp \VEC{\hat{c}}$, and the results are presented in Fig.~\ref{fig:rawdata2} and Fig.~\ref{fig:rawdata3}.
The AFM order is characterized by magnetic reflections arising as satellites to the nuclear Bragg reflections with modulation vector $(1/3, 1/3, k_z)$, where $k_z$ changes as a function of the applied field direction and magnitude.

The evolution of the magnetic order at $T = 300$\,mK for $\VEC H \parallel \VEC{\hat{c}}$ is depicted in Fig.~\ref{fig:rawdata2}(a) where for critical fields $H > H_{c_1}^{\parallel}$ the system shows commensurate ordering.
Three different phases can be distinguished up to the maximum investigated magnetic field of 3.8\,T: (i) for $H_{c_1}^{\parallel} \leq H < H_{c_2}^{\parallel}$ the magnetic wave vector slightly shifts and becomes $(1/3, 1/3, 1/6)$ with a second less intense magnetic peak appearing at $(1/3, 1/3, 1/3)$, (ii) for $H_{c_2}^{\parallel} \leq H \leq H_{c_3}^{\parallel}$ the system enters into the 1/3 plateau phase~\cite{Kim_2022} and this is accompanied by a first sudden change in $\VEC k$ which is now located at $(1/3, 1/3, 0)$ [up-up-down (UUD) phase], and (iii) a second abrupt jump results to $\VEC k = (1/3, 1/3, 1/3)$ for $H_{c_3}^{\parallel} < H \leq H_{\text{max}}^{\parallel}$.
In contrast, at $T = 1200$\,mK for $H_{c_1}^{\parallel} \leq H < H_{c_2}^{\parallel}$, a continuous diminution of the magnetic peak intensity and a shift of $\VEC k$ towards the K-point are observed (see Fig.~\ref{fig:rawdata2}(c)).
When the system is tuned into the UUD phase, a strong enhancement of the peak intensity is detected before reaching the spin-polarized (SP) state.

Fig.~\ref{fig:rawdata2}(b) shows the reciprocal space scans at $\VEC Q = (1/3, 1/3, Q_{l})$ for $\VEC H \perp \VEC{\hat{c}}$ at $T = 300$\,mK.
Based on the scattering geometry given in Section~\ref{sec:methods} the magnetic field points along the $[1\bar{1}0]$ direction of the trigonal symmetry.
With increasing magnetic field we observed the following: (i) in the narrow range $H_{c_1}^{\perp} \leq H < H_{c_2}^{\perp}$ the zero-field AFM1 might be co-existing with another phase that can be indexed with $\VEC k = (1/3, 1/3, 1/12)$, (ii) for $H_{c_2}^{\perp} \leq H < H_{c_3}^{\perp}$ the $(1/3, 1/3, 1/12)$ reflection becomes very sharp, while three additional less intense peaks appear in commensurate positions with $k_z'=1/6$, 1/3 and 5/12, (iii) for $H_{c_3}^{\perp} \leq H < H_{c_4}^{\perp}$ the magnetic peak is located at $(1/3, 1/3, 1/6)$ and a second weaker at $(1/3, 1/3, 1/3)$ and, (iv) for $H_{c_4}^{\perp} \leq H < H_{\text{SP}}^{\perp}$ the system becomes incommensurate with $k_z$ changing continuously with field, before reaching the spin-polarized state. 
It is worth mentioning that at $T = 1200$\,mK no field-induced phase transition is detected until the system enters the SP state [see Fig.~\ref{fig:rawdata2}(d)].
Representative $\VEC Q$ scans for every phase showing the main and secondary magnetic peak positions for the two field directions are summarized in Fig.~\ref{fig:rawdata3}.

\subsection{Temperature and magnetic field phase diagrams}

\begin{figure}[ht]
    \includegraphics[width=\columnwidth]{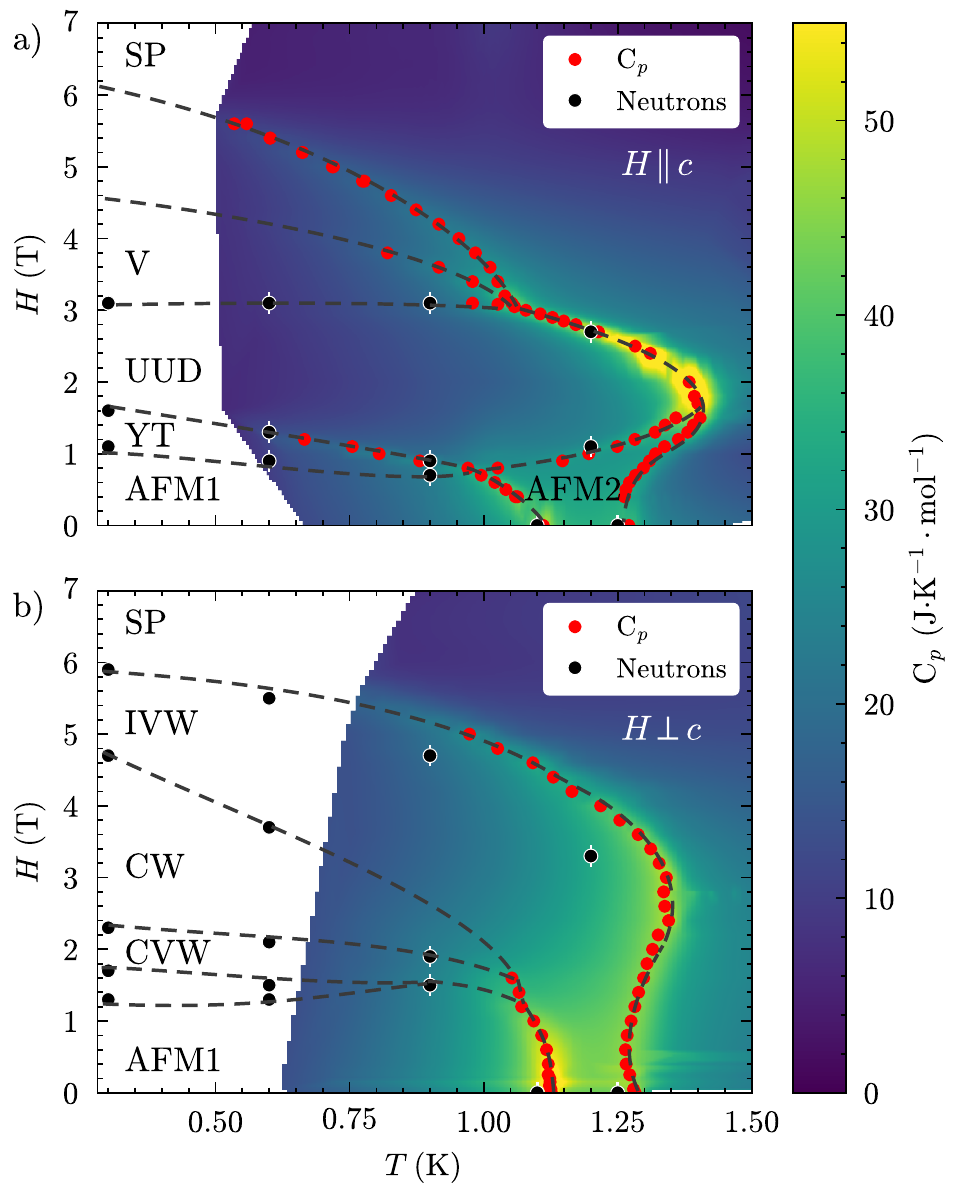}
    \caption
    {Temperature and magnetic field phase diagrams of Na$_2$BaMn(PO$_4$)$_2$ with magnetic field applied along (a) the $c$-axis and (b) in the $ab$-plane. Red and black circles indicate critical fields determined from specific heat (background heat map) and neutron diffraction measurements, respectively. Phase boundaries are marked by dashed gray lines and are guides for the eyes. }
    \label{fig:phasediag}
\end{figure}

From the peak positions in the heat capacity data and the changes of the propagation vector we constructed the temperature and magnetic field phase diagrams shown in Figure~\ref{fig:phasediag}.
The phase boundaries for the spin polarized state for the two field directions are in agreement with a previous report~\cite{Kim_2022}; however, our data suggest more complex $H-T$ phase diagrams.
For $\VEC H \parallel \VEC{\hat{c}}$, we evidence the formation of two pocket phases; a narrow between the AFM1 and the UUD phase, and AFM2.
For $\VEC H \perp \VEC{\hat{c}}$ there is an absence of a phase transition for AFM2, a narrow region for $T < T_{N_1}$ where the AFM1 incommensurate ground state coexists with a commensurate phase, and a phase transition between a commensurate and incommensurate phase at higher magnetic fields.

The existence of a 1/3 plateau (UUD phase) is reported also in other classical TLA systems with $S = 5/2$, e.g. Rb$_4$Mn(MoO$_4$)$_3$~\cite{Ishii_2011}, Ba$_3$MnNb$_2$O$_9$~\cite{Lee_2014} (Mn$^{+2}$) and RbFe(MoO$_4$)$_2$~\cite{Svistov_2006} (Fe$^{+3}$).
For classical systems the stability of the UUD phase in a finite temperature range indicates that thermal fluctuations are lifting the degeneracy of the ground state~\cite{Villain_1980,Henley1989}; otherwise, their absence would result in the appearance of the UUD phase at a single point. 
Thermal fluctuations are responsible for lowering the free energy of the system by selecting the highest entropic state, where two spins are aligned parallel and the other spin antiparallel to the direction of the external applied magnetic field.
As the magnetic field further increases, a canted version of the UUD phase becomes stable (umbrella or V phase)~\cite{Seabra_2011,Yamamoto_2014}.

\subsection{Spin structures under magnetic field}
\label{sec:spin-structures}

The spin structure models of Na$_2$BaMn(PO$_4$)$_2$ are based on the representation analysis of the modulation vector $\VEC k = (1/3, 1/3, k_z)$, with the component $k_z$ varying throughout the phase diagram, as summarized in Section~\ref{sec:phase-transitions}. 
For all values of $k_z$, there are the same three irreducible representations (irreps) of the magnetic propagation vector group $\Gamma_\mathrm{mag} = \Gamma_1 \oplus \Gamma_2 \oplus \Gamma_3$, which are presented in Table~\ref{tab:irreps}.
$\Gamma_1$ mode is a collinear arrangement of moments along the $c$-axis, with the up-up-down (UUD) amplitude pattern for zero phase, and plus-minus-zero for phase $\pi/4$.
The modulation along the $c$-axis direction given by $k_z$ modulates the spin arrangement between these two patterns.
$\Gamma_2$ has magnetic moments coplanar in the (001) plane, with moments forming the 120$^\circ$ structure along Mn--Mn bonds, and the $k_z$ modulation rotates the moments around the $c$-axis, showing right-handed helical modulation.
$\Gamma_3$ is the same as $\Gamma_2$, but the $k_z$ modulation is left-handed. 
An explicit mode decomposition and notation are given in the Supplementary Material (see Eq.~S1~\cite{Supplemental}).

\begin{table}[b]
\centering
\caption{Irreducible representations of the magnetic moments of Mn ions in Na$_2$BaMn(PO$_4$)$_2$. Irreps are derived for the Wyckoff position $1a$ in the space group $P\bar{3}$ and modulation vector $\VEC k = (1/3, 1/3, k_z)$. Columns contain irreps symbols, normalized basis vectors and description of the spin structures.}
\label{tab:irreps}
\begin{tabular}{l l l}
\hline\hline
    irreps & basis vector & description  \\ \hline
    
    $\Gamma_1$ & [0 0 1] & collinear with the $c$-axis \\
     &  & modulating between UUD and +-0 \\ [3pt]
     
    $\Gamma_2$ & [0.612 0 0] + & 120\deg\ in $ab$-plane, right-handed \\
     & - $i$[0.354 0.707 0] & helical modulation along the $c$-axis \\ [3pt]
        
    $\Gamma_3$ & [0.612 0 0]+ & 120\deg\ in $ab$-plane, left-handed \\
     & + $i$[0.354 0.707 0] & helical modulation along the $c$-axis \\
\hline\hline
\end{tabular}
\end{table}

Before presenting the results there are three important points to address regarding our measurements.
First, unpolarized neutron diffraction can not distinguish between opposing chiralities of the magnetic structure, that is the difference between $\Gamma_2$ and $\Gamma_3$, but can determine the ratio of these mode amplitudes.
Therefore, the magnetic structure can be described with reference to the in-plane component ($\Gamma_2$ and $\Gamma_3$ mixing) and out-of-plane component ($\Gamma_1$).
Second, neutron diffraction cannot determine the global phase of the magnetic modulation, which is crucial to accurately depict the structure with a commensurate modulation vector.
We have chosen the phase to best match the theoretical models presented in the next section.
Finally, it is important to note that the refinement of the satellite reflection intensities allowed us to determine the antiferromagnetic component of the magnetic order.
Under an applied magnetic field, there is an additional ferromagnetic component arising from magnetization that contributes to the nuclear reflections, however, determining its precise value from neutron diffraction measurements is unreliable.
For depicting the magnetic structures under field we used the  magnetization values determined in Ref.~\cite{Kim_2022}, also shown in Fig.~\ref{fig:anisotropy}.
A comprehensive list of the refinement parameters, including mode amplitudes, ferromagnetic components, and global phases, is given in the Supplementary Material (see Table~SII, and goodness-of-fit plots in Fig.~S2~\cite{Supplemental}).

\begin{figure}[!th]
    \includegraphics[width=\columnwidth]{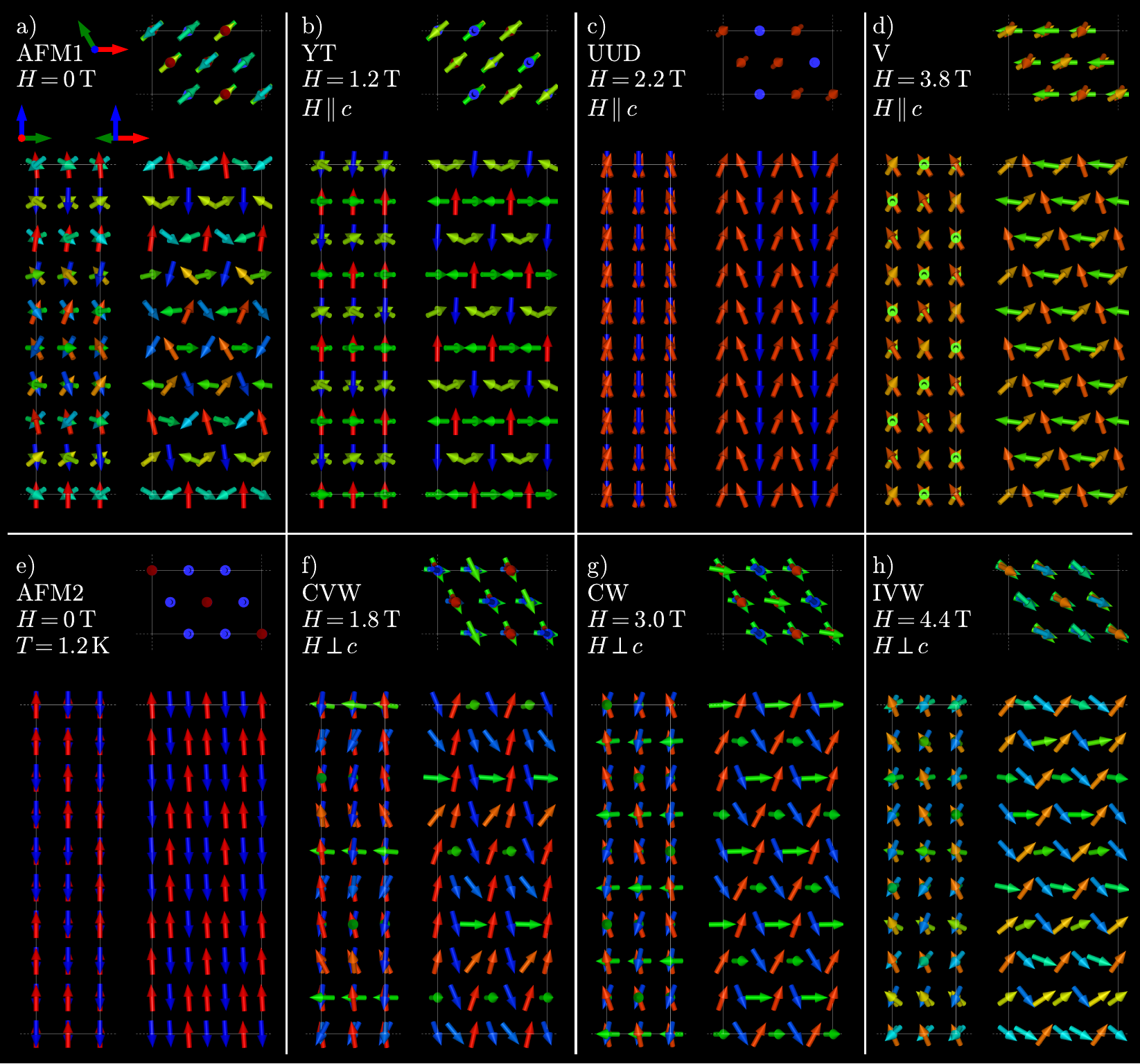}
    \caption
    {Magnetic structures of Na$_2$BaMn(PO$_4$)$_2$ based on single crystal neutron diffraction measurements under applied field. 
    All structures are determined at 600\,mK except the AFM2 phase (e). 
    Each structure is depicted from the directions along the $c$, $a$, and $a^{*}$-axis. 
    Axes coordinates for these views are depicted only in (a) showing $a_1$, $a_2$ and $c$-axis in red, green, blue respectively.
    (a) At zero field and 600\,mK, the spin configuration is incommensurate, co-planar with Y arrangement of spins and components along the $c$-axis and $[120]$ direction.
    (e) At 1200\,mK it is collinear along the $c$-axis, with moments in an up-up-down pattern.
    When a magnetic field is applied along the $c$-axis, spins gradually align with the field forming (b) alternating YT layers at 1.2~T, (c) the UUD phase at 2.2~T, and (d) the umbrella V arrangement at 3.8~T.
    For the field applied along the $[1 \bar{1} 0]$ direction the spins also gradually align along the field, with (f) commensurately alternating V-W layers at 1.8\,T, (g) commensurately stacked W layers at 3\,T, and (h) incommensurately modulated V-W layers at 4.4\,T.
    A detailed description of all phases is given in the main text; numerical values used to render all panels are listed in the Supplementary Material in Table~SII~\cite{Supplemental}.
    }
    \label{fig:spin-structures}
\end{figure}

The spin arrangement at 600~mK and zero-field (AFM1 phase) shows in-plane and out-of-plane components, as displayed in Figure~\ref{fig:spin-structures}(a).
The in-plane modes $\Gamma_2$ and $\Gamma_3$ sum up to a collinear component along the $[120]$ direction, which in combination with the out-of-plane mode $\Gamma_1$ results in a coplanar structure with Y arrangement of spin directions in the $z=0$ layer.
The incommensurability of the modulation vector changes the phase of the spin configuration along the $[001]$ direction, resulting in cycloidal modulation of Mn spins with $\approx$60\deg\ interlayer phase shift and effective rotation of the Y configuration along the $c$-axis direction.
The average magnitude of the individual moments is $3.70(8)\,\mu_\mathrm{B}$.
Magnetic ordering at 1200\,mK and zero field (AFM2 phase) is pure $\Gamma_1$ mode, that is, a collinear UUD structure along the $c$-axis, as shown in Figure~\ref{fig:spin-structures}(e).
The absolute value of the magnetic moment is reduced to $\mu=2.07(1) \mu_\mathrm{B}$ and the structure resembles the 600\,mK arrangement dynamically disordered within the $ab$-plane.
Our refinements in zero-field match those reported in Ref.~\cite{Zhang_2024}, where the authors determined the magnetic structures based on neutron powder diffraction of Na$_2$BaMn(PO$_4$)$_2$ at 67\,mK and 1.25\,K, with small differences in the modulation vector.

The field applied along the $c$-axis breaks the incommensurate modulation of the spin arrangement and introduces intermediate, commensurate phases before field-polarizing the spins [see Figures~\ref{fig:spin-structures}(b)-(d)].
At 1.2~T (YT phase) the refinement resembles the ground state Y configuration with spins slightly rotated along the field direction.
The modulation along the $c$-axis alternates between the distorted Y configuration and T configuration, where the latter depicts two spins arranged antiferromagnetically within the $ab$-plane and one spin along the $c$-axis.
Note, that the refinement at 1.2~T was poor compared to the other refinements (see Fig.~S2(b) in Supplementary Material~\cite{Supplemental}), and should be treated as a rough approximation.
Next, at 2.2\,T spins arrange in the UUD configuration along the $c$-axis with a small in-plane component (UUD phase). 
The in-plane component is collinear with the $[120]$ direction, and five times smaller than the $c$-axis component.
Finally, at 3.8\,T the spins rearrange to an umbrella configuration (V phase), where all spins have positive components along the field, while the in-plane component spreads them around in a 120\deg\ configuration.

With the field applied in-plane, all measured phases show almost coplanar arrangement, as the dominant component is $\Gamma_1$ mode with $c$-axis component, and uniform magnetization aligning the moments toward the $[1 \bar{1} 0]$ direction.
As a consequence, the spins are lying within the $(110)$ plane, that is, they have dominating components along the $c$-axis and field direction [see Figures~\ref{fig:spin-structures}(f)-(h)].
At 1.8\,T the spins arrange in alternating V and W layers, with opening along the applied field direction $[1 \bar{1} 0]$ (CVW phase).
The V configuration is a superposition of UUD along the $c$-axis and uniform magnetization along the applied field resulting in spins out-of-plane, while the W configuration describes one spin along the field direction and the other two spread up and down with respect to the first one.
When increasing the field to 3\,T there are only W layers with commensurate modulation along the $c$-axis and opening of $\approx$120\deg (CW phase). 
At 4.4\,T the structure is incommensurate again and modulates between the W and V configuration (IVW phase) with opening of $\approx$85\deg. 
The latter incommensurate configuration is also referred to as a fan phase~\cite{Nagamiya_1968}.

\subsection{Spin model Hamiltonian}

\begin{figure}[h]
    \setlength{\unitlength}{0.1\textwidth}
    \includegraphics[width=0.48\textwidth]{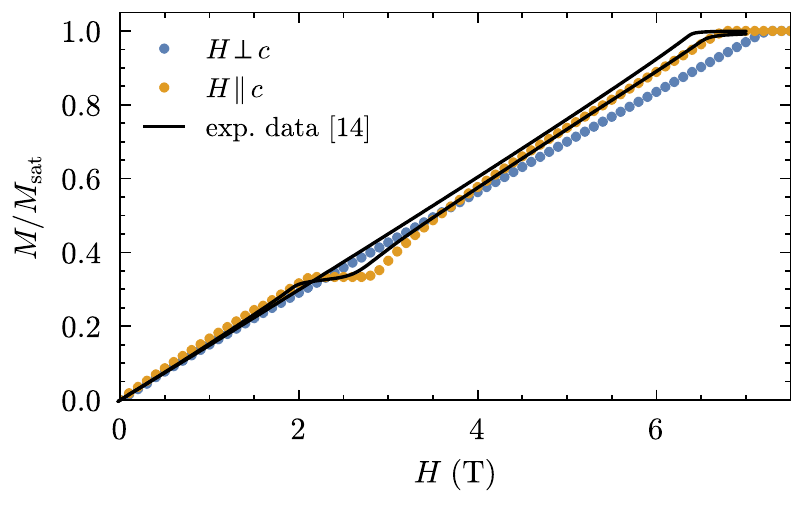}
    \caption
    {Magnetization as a function of applied fields along and perpendicular to the easy-axis. Experimental data based on Ref.~\cite{Kim_2022}. Simulation (color circles) including single-ion anisotropy with easy-axis along $c$. The length of the plateau is proportional to the single-ion anisotropy. }
    \label{fig:anisotropy}
\end{figure}

The localized magnetic moments in Na$_2$BaMn(PO$_4$)$_2$ form a triangular lattice in the $ab$-plane, with a Mn--Mn stacking along the $c$-axis as described in Section~\ref{sec:spin-structures}.
The phase diagrams and the spin structure discussed previously indicate the dynamics of a frustrated antiferromagnetic triangular lattice~\cite{Seabra_2011,Kim_2022}.
Initially, for Na$_2$BaMn(PO$_4$)$_2$ we considered a two-dimensional (2D) nearest-neighbor-only classical Heisenberg model with $J_1=-0.11$\,meV and $S=5/2$.
The significant single-ion anisotropy of Mn$^{+2}$ ions points to an easy-axis anisotropy in the system.
In order to reproduce the plateau in the reported magnetization measurements as a function of applied fields along the $c$-axis~\cite{Kim_2022}, $k^c=0.035$\,meV was added to our model Hamiltonian (see Fig.~\ref{fig:anisotropy}). 
Our first attempt to reproduce the experimental results with this simple model is justified by the fact that two successive magnetic transitions at zero magnetic field are reported in other 2D triangular-lattice easy-axis AFM with high spin number~\cite{Kitazawa_1999,Kimura_2008, Ishii_2011}.

\begin{figure}[h]
    \setlength{\unitlength}{0.1\textwidth}
    \includegraphics[width=0.48\textwidth]{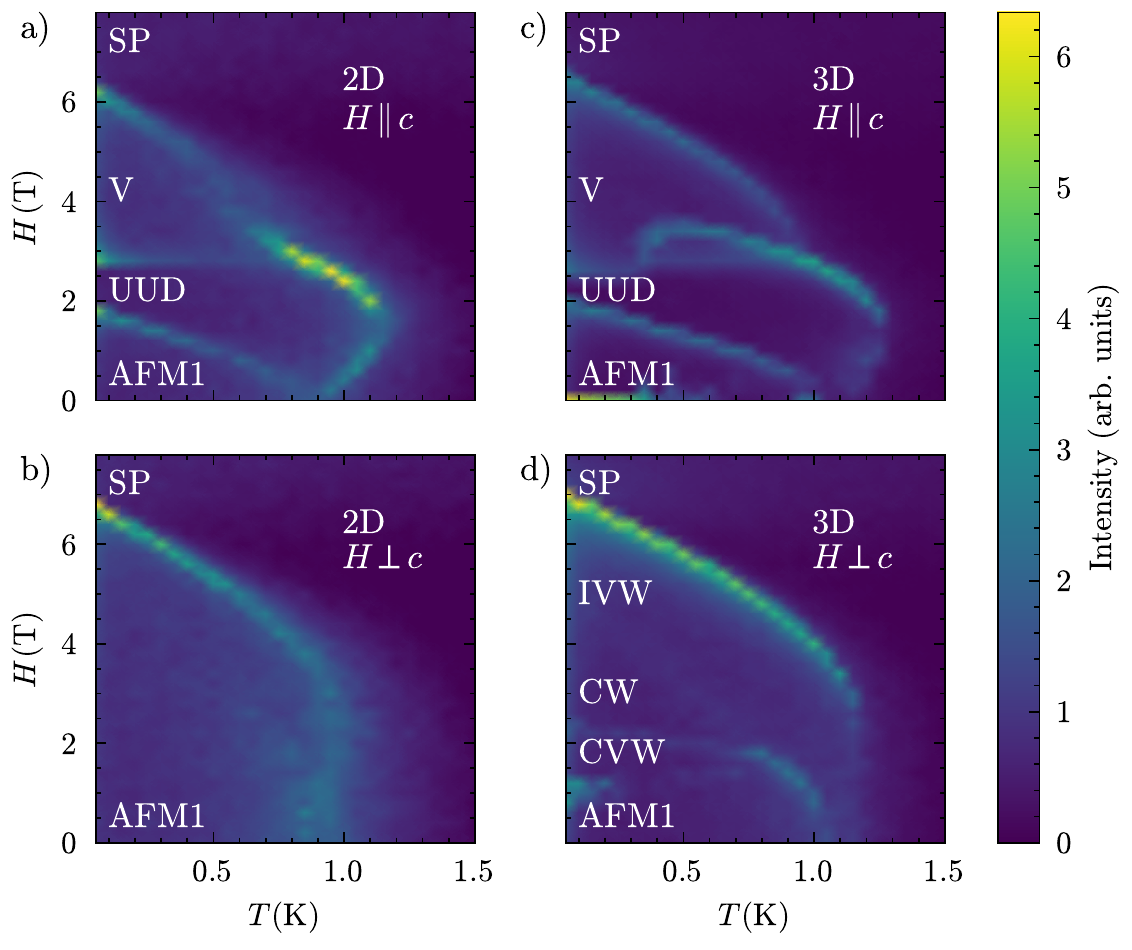}
    \caption
    {Color maps of the calculated specific heat as a function of applied magnetic field and temperature for the (a), (b) 2D model and (c), (d) 3D model.}
    \label{fig:specif_heat_two_models}
\end{figure}

\begin{figure}[h]
    \setlength{\unitlength}{0.1\textwidth}
    \includegraphics[width=0.48\textwidth]{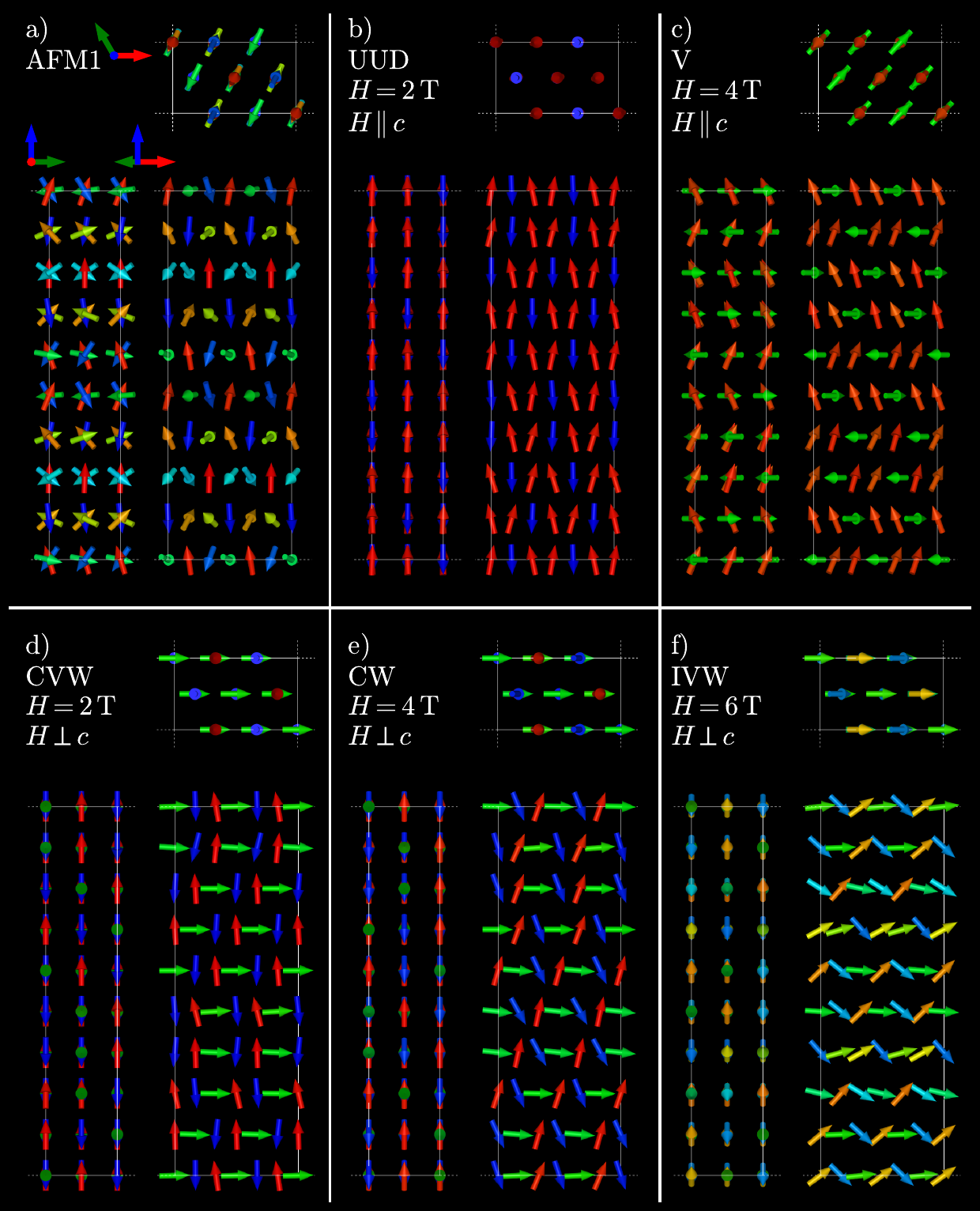}
    \caption
    {The spin arrangements in the different phases based on the 3D model at $T=0$\,K for: (a) the ground state, (b)-(c) for external field applied out-of-plane, i.e., along the $c$-axis, and (d-f) in-plane, i.e., along the $a$-axis.}
    \label{fig:model_spins}
\end{figure}

The Monte Carlo simulations with this 2D model resulted in the color maps of the specific heat as a function of temperature and applied field which are shown in Figs~\ref{fig:specif_heat_two_models}(a) and \ref{fig:specif_heat_two_models}(b).
These heat maps reveal the boundaries between the different phases of the model.
At finite temperatures and in zero field, a temperature separation of $\sim$0.05\,K is observed between the AFM1 and AFM2 phases, and the system becomes paramagnetic at $\sim$0.95\,K. 
When the field is applied along the easy-axis $c$, we identify at least four regions of the phase-space diagram in Fig.~\ref{fig:specif_heat_two_models}(a).
Through spin dynamics simulations, we determined the spin configuration in each phase at zero temperature.
AFM1: The ground state at zero field corresponds to a noncollinear $120^\circ$ spin configuration where one of the three spins in the magnetic unit cell points along the easy-axis (out-of-plane).
UUD: At 2\,T, the system displays a collinear up-up-down spin configuration.
V: This phase resembles the UUD state, although it is noncollinear, and it is marked by the rotations of the down spin towards the direction of the applied field.
SP: At 6\,T, the Zeeman energy dominates the exchange interactions and spin polarizes the system.
In contrast, when the field is applied in-plane, only two regions (AFM1 and SP) are distinguishable through the specific heat color map, as shown in Fig.~\ref{fig:specif_heat_two_models}(b).
Starting from the ground state at zero field, when the external field dominates the single-ion anisotropy, one spin starts to point parallel to the field.
At 3\,T, we observe that two spins are perpendicular to the field, antiparallel to each other, but aligned along the easy-axis, while the third spin remains parallel to the applied field.
For higher fields, all spins continuously align with the field until the spin-polarized state is reached, slightly below 7\,T.
This phase sequence is consistent with the classical easy-axis triangular-lattice Heisenberg model~\cite{Seabra_2011}.

This simple 2D model already seems to capture some features and the overall shape of the experimentally determined phase diagrams.
It is important to mention that the experimental phase diagrams shown in Figure~\ref{fig:phasediag} were obtained from a combination of neutron diffraction and specific heat measurements, while the theoretical phase diagrams were determined from the specific heat results from Monte Carlo simulations.
Therefore, it is possible that the specific heat will not reveal all the field-induced transitions between the different phases.
In addition, neutron diffraction data suggest that the magnetic order is characterized by an incommensurate wave-vector along the $c$-axis, as shown in Figs.~\ref{fig:rawdata2} and~\ref{fig:rawdata3}, indicating that Na$_2$BaMn(PO$_4$)$_2$ is a three-dimensional (3D) easy-axis antiferromagnet with considerable interlayer couplings.
This corresponds to a spin configuration based on the $120^\circ$ state, which offers a global rotation from one layer to the next.
It is also worth mentioning that based on Fig.~\ref{fig:rawdata3}(b), at zero field intense magnetic Bragg peaks are observed at positions with magnetic propagation vectors $\VEC k = (1/3, 1/3, \pm k_z)$.
Propagation vectors of this form have been attributed to frustrated interlayer couplings and are reported in compounds with stacked magnetic triangular layers such as the isostructural RbFe(MoO$_4$)$_2$~\cite{Hearmon_2012}. 

In order to include the interaction between layers in our model, we attempted to reproduce the spiraling spin configuration in Na$_2$BaMn(PO$_4$)$_2$ via a frustration induced by two interlayer couplings.
While the frustration 3D-coupling model was recently proposed for Na$_2$BaCo(PO$_4$)$_2$~\cite{Woodland_2025}, our study is the first to apply it to Na$_2$BaMn(PO$_4$)$_2$, demonstrating its relevance also for S = 5/2 systems.
It is based on the competition between $J_2$, $J_a$, and $J_b$, see Fig.~\ref{fig:crystal-structure}(d).
According to this model, the propagation vector along $c$ is given by~\cite{Woodland_2025}:
\begin{equation}
    \tan(2\pi k_z) = \frac{3\sqrt{3}(J_b-J_a)}{3(J_b+J_a) - 2J_2}.
\end{equation}
In the $P\bar3m1$ structure, $J_a$ and $J_b$ are equal by symmetry; however, if the system's crystal structure is $P\bar3$, thus losing a vertical mirror symmetry, $J_a$ and $J_b$ become independent of each other.
The competition between pairs $J_a$ and $J_b$, or $J_{a(b)}$ and $J_2$ is enough to form the observed spin spiral due to frustration.
Our simulations indicate that the $J_a$-$J_2$ model results in a better description of the experimental phase diagrams presented in  Fig.~\ref{fig:phasediag}.
We determined that $J_a = -0.0013$ and $J_2 = -0.002$\,meV (1-2\% of $J_1$) reproduce a propagation vector and roughly the phase boundaries of the field-temperature phase diagrams.

The resulting ground state spin configuration matches very well the experimentally determined one, with the exception that the model has a commensurable wavelength of 5 layers to allow for technical simplifications of the simulations.
The specific-heat phase diagram for field out-of-plane is shown in Fig.~\ref{fig:specif_heat_two_models}(c), which is very similar to the 2D case, except for a new pocket created between the UUD and V phases.
We note that at 2\,T, with the 3D coupling, the spin configuration is similar to the up-up-down state.
For a field in-plane, while the 2D model only featured a single ordered phase, the 3D coupling introduced several phase boundaries to the diagram, although some of them are very faint, as seen in Fig.~\ref{fig:specif_heat_two_models}\,(d).
It is worth mentioning that the 3D model results in $\sim$0.18\,K. separation between the two successive zero-field magnetic transitions, and the paramagnetic state is observed at $\sim$1.18\,K. 
The evolution of the spin configurations for the two applied field directions at zero temperature is also displayed in Figure~\ref{fig:model_spins}, and a further comparison between the simulated and experimentally determined magnetic structures is presented in detail in the Supplementary Material (see Fig.~S3~\cite{Supplemental}).
At this stage a perfect agreement between the experimentally determined magnetic structures and the theoretically obtained ones cannot be achieved, however, several features are captured by the proposed 3D model.

Detailed inelastic neutron scattering studies on single crystals for both field directions and additional theoretical modeling would be crucial to determine the Hamiltonian of the Na$_2$BaMn(PO$_4$)$_2$ system. 
Such measurements have been used to determine the microscopic Hamiltonian in the analogue RbFe(MoO$_4$)$_2$~\cite{White_2013}, which also exhibits Y, UUD, and V phases and field-induced incommensurate stacking along the $c$-axis~\cite{Sakhratov_2022}.
For example, an alternative scenario might involve an Ising-like XXZ Hamiltonian where (i) the width of the 1/3 UUD plateau in Fig.~\ref{fig:phasediag}(a) will depend on the XXZ exchange anisotropy of the $J_1$ bond, and (ii) the separation between the two zero-field transitions reported at 1.13 and 1.28\,K will rely on the XXZ nature and on the 3D couplings~\cite{Jia_2024}.

Our reported $H-T$ phase diagrams in Na$_2$BaMn(PO$_4$)$_2$ largely follow the archetypal frustrated triangular-lattice antiferromagnet scenario (coplanar Y and V phases separated by the UUD state) as summarized in Ref.~\cite{Starykh_2015}. 
A key difference from the strictly 2D picture is that the propagation vector acquires a field-tunable out-of-plane component, $k_z(H,T)\neq 0$, which signals weak and frustrated interlayer exchange and is captured by our minimal 3D model. 
We also note that the very low-field region of the TLA phase diagram is explicitly flagged as unsettled in Ref.~\cite{Starykh_2015} and is where we capture the AFM2 phase. 
Its character, i.e., a longitudinal (easy-axis) ordering that precedes the coplanar AFM1 on cooling, is consistent with the XXZ triangular model in the presence of weak easy-axis anisotropy and weak 3D coupling, and with theoretical expectations for zero field~\cite{Kawamura_2010}. 
Within this context, RbFe(MoO$_4$)$_2$ realizes the canonical Y--UUD--V sequence with commensurate stacking for field applied in-plane~\cite{Sakhratov_2022}, whereas in CuCrO$_2$ weak interlayer couplings and easy-axis anisotropy drive incommensurate stacking and a rich field-induced cascade~\cite{Mun_2014,Sakhratov_2016}; see also Ref.~\cite{Frontzek_2011} for the crucial role of interlayer exchange in the incommensurate phases.
In Na$_2$BaMn(PO$_4$)$_2$, the combination of easy-axis anisotropy and frustrated interlayer exchange could explain both the evolution $k_z\neq 0$ in applied field and the finite zero-field AFM2 pocket.

\section{Conclusions}

To conclude, we have reported detailed heat capacity, X-ray, and neutron diffraction measurements in single crystals of the $S=5/2$ triangular-lattice antiferromagnet Na$_2$BaMn(PO$_4$)$_2$.
Based on X-ray diffraction refinements, the crystal structure exhibits subtle distortions that break the vertical mirror planes of the trigonal symmetry and reduce the space group from $P\bar{3}m1$ to $P\bar3$.
Heat capacity and neutron diffraction measurements under magnetic fields applied along the $c$-axis and in the $ab$-plane allowed us to identify several field-induced transitions in the temperature and magnetic field phase diagrams.
We refined the magnetic structures of Na$_2$BaMn(PO$_4$)$_2$ according to the representation analysis of the modulation vector $\VEC k = (1/3, 1/3, k_z)$, where the component $k_z$ changes with magnetic field.
In the ground state, the out-of-plane incommensurate component $k_z$ of the propagation vector indicates non-negligible interlayer couplings.
By employing a classical Heisenberg Hamiltonian that includes a single-ion anisotropy term ($k^c=0.035$\,meV), nearest-neighbors interactions in plane ($J_1 = -0.11$\,meV), and interlayer couplings ($J_a = -0.0013$ and $J_2 = -0.002$\,meV), and in combination with Monte Carlo simulations, we reproduced several features in the reported phase diagrams and the spin structures.
Our data and simulations show that Na$_2$BaMn(PO$_4$)$_2$ is far more three-dimensionally coupled than assumed, and that a minimal frustrated interlayer Heisenberg model captures the main features of its rich field-induced phase diagram.

\section{Data Availability}

The neutron data collected at the ILL are available at Refs.~\onlinecite{data_D23,data_D23b}. 
The data treatment and figure reproduction workflow (notebook, scripts, and raw data inputs) are available on Figshare~\cite{data}.

\section{Acknowledgments}

We acknowledge help from N. Qureshi for Mag2POL and from O. Fabelo for the initial X-ray single crystal diffraction measurements performed at ILL.
We thank B. Normand and A. M. L\"auchli for discussions and comments.
This work was supported by the Czech Science Foundation GA\v CR under the Junior Star Grant No. 21-24965M (MaMBA), by the Czech Ministry of Education Youth and Sports Grant No. LUABA24056 (BaCQuERel), and by the NCCR MARVEL, a National Centre of Competence in Research, funded by the Swiss
National Science Foundation (Grant No. 205602).
Crystals were grown and characterized in MGML (mgml.eu), which is supported within the program of Czech Research Infrastructures (project No. LM2023065).

\bibliography{main}

@article{Woodland_2025,
  title = {{From continuum excitations to sharp magnons via transverse magnetic field in the spin-$\frac{1}{2}$ Ising-like triangular lattice antiferromagnet Na$_2$BaCo(PO$_4$)$_2$}},
  author = {Woodland, Leonie and Okuma, Ryutaro and Stewart, J. Ross and Balz, Christian and Coldea, Radu},
  journal = {Phys. Rev. B},
  volume = {112},
  issue = {10},
  pages = {104413},
  numpages = {21},
  year = {2025},
  doi = {10.1103/1pvl-kzjm},
  url = {https://link.aps.org/doi/10.1103/1pvl-kzjm}
}

@incollection{Nagamiya_1968,
title = {{Helical Spin Ordering—1 Theory of Helical Spin Configurations}},
editor = {Frederick Seitz and David Turnbull and Henry Ehrenreich},
series = {Solid State Physics},
publisher = {Academic Press},
volume = {20},
pages = {305-411},
year = {1968},
issn = {0081-1947},
doi = {https://doi.org/10.1016/S0081-1947(08)60220-9},
url = {https://www.sciencedirect.com/science/article/pii/S0081194708602209},
author = {Takeo Nagamiya},
}

@article{D23-LSF,
	title = {{The new thermal-neutron diffractometer D23}},
	journal = {{ILL Technical Report}},
	year = {1999},
    volume={123},
    pages={92–93},
	author = {E. Ressouche and J. Chiapusio and B. Longuet and F. Mantegazza and J. Flouquet},
    url={https://inis.iaea.org/records/cmv46-zfe87}
}

@article{shelx,
  title={{A short history of SHELX}},
  author={Sheldrick, George M},
  journal={Acta Crystallographica Section A: Foundations of Crystallography},
  volume={64},
  number={1},
  pages={112--122},
  year={2008},
  publisher={International Union of Crystallography},
  doi={10.1107/S0108767307043930},
  url={https://doi.org/10.1107/S0108767307043930}
}

@misc{Crysalis,
	author = {Agilent}, 
	title = {{CrysAlis$^{Pro}$ Software System}},
	howpublished = {ver. 1.171.36.28, Agilent Technologies UK Ltd., Oxford, UK},
	year = 2013}

@article{muller_spirit:_2019,
	title = {Spirit: {Multifunctional} framework for atomistic spin simulations},
	author = {M\"uller, Gideon P. and Hoffmann, Markus and Di\ss{}elkamp, Constantin and Sch\"urhoff, Daniel and Mavros, Stefanos and Sallermann, Moritz and Kiselev, Nikolai S. and J\'onsson, Hannes and Bl\"ugel, Stefan},
    journal = {Phys. Rev. B},
    volume = {99},
    issue = {22},
    pages = {224414},
    numpages = {16},
    year = {2019},
    doi = {10.1103/PhysRevB.99.224414},
    url = {https://link.aps.org/doi/10.1103/PhysRevB.99.224414}
}

@article{evans_atomistic_2014,
	title = {Atomistic spin model simulations of magnetic nanomaterials},
	volume = {26},
	issn = {0953-8984},
	number = {10},
	journal = {Journal of Physics: Condensed Matter},
	author = {Evans, R. F. L. and Fan, W. J. and Chureemart, P. and Ostler, T. A. and Ellis, M. O. A. and Chantrell, R. W.},
	year = {2014},
	pages = {103202},
	url = {https://dx.doi.org/10.1088/0953-8984/26/10/103202},
	doi = {10.1088/0953-8984/26/10/103202}
}

@misc{vampire,
      note = {VAMPIRE software package version 6.0.0 available from \url{https://vampire.york.ac.uk}
(Version 4c9651daecc86c3e6b6df0c8055fcd9e89fd900c).}
}

@article{Savary_2017,
doi = {10.1088/0034-4885/80/1/016502},
url = {https://dx.doi.org/10.1088/0034-4885/80/1/016502},
year = {2016},
month = {nov},
publisher = {IOP Publishing},
volume = {80},
number = {1},
pages = {016502},
author = {Savary, Lucile and Balents, Leon},
title = {Quantum spin liquids: a review},
journal = {Reports on Progress in Physics}
}

@article{Broholm_2020,
author = {C. Broholm  and R. J. Cava  and S. A. Kivelson  and D. G. Nocera  and M. R. Norman  and T. Senthil },
title = {Quantum spin liquids},
journal = {Science},
volume = {367},
number = {6475},
pages = {eaay0668},
year = {2020},
doi = {10.1126/science.aay0668},
URL = {https://www.science.org/doi/abs/10.1126/science.aay0668}
}

@article{Balents_2010,
  title = {Spin liquids in frustrated magnets},
  volume = {464},
  ISSN = {1476-4687},
  url = {http://dx.doi.org/10.1038/nature08917},
  DOI = {10.1038/nature08917},
  number = {7286},
  journal = {Nature},
  publisher = {Springer Science and Business Media LLC},
  author = {Balents,  Leon},
  year = {2010},
  pages = {199–208}
}

@article{Wen_1991,
  title = {Mean-field theory of spin-liquid states with finite energy gap and topological orders},
  author = {Wen, X. G.},
  journal = {Phys. Rev. B},
  volume = {44},
  issue = {6},
  pages = {2664--2672},
  numpages = {0},
  year = {1991},
  doi = {10.1103/PhysRevB.44.2664},
  url = {https://link.aps.org/doi/10.1103/PhysRevB.44.2664}
}

@article{Seabra_2016,
  title = {Novel phases in a square-lattice frustrated ferromagnet: $\frac{1}{3}$-magnetization plateau, helicoidal spin liquid, and vortex crystal},
  author = {Seabra, Luis and Sindzingre, Philippe and Momoi, Tsutomu and Shannon, Nic},
  journal = {Phys. Rev. B},
  volume = {93},
  issue = {8},
  pages = {085132},
  numpages = {23},
  year = {2016},
  doi = {10.1103/PhysRevB.93.085132},
  url = {https://link.aps.org/doi/10.1103/PhysRevB.93.085132}
}

@article{Lee_2014,
  title = {Magnetic phase diagram and multiferroicity of {Ba$_3$MnNb$_2$O$_9$}: A spin-$\frac{5}{2}$ triangular lattice antiferromagnet with weak easy-axis anisotropy},
  author = {Lee, M. and Choi, E. S. and Huang, X. and Ma, J. and Dela Cruz, C. R. and Matsuda, M. and Tian, W. and Dun, Z. L. and Dong, S. and Zhou, H. D.},
  journal = {Phys. Rev. B},
  volume = {90},
  issue = {22},
  pages = {224402},
  numpages = {8},
  year = {2014},
  doi = {10.1103/PhysRevB.90.224402},
  url = {https://link.aps.org/doi/10.1103/PhysRevB.90.224402}
}

@article{Rosales_2013,
  title = {Broken discrete symmetries in a frustrated honeycomb antiferromagnet},
  author = {Rosales, H. D. and Cabra, D. C. and Lamas, C. A. and Pujol, P. and Zhitomirsky, M. E.},
  journal = {Phys. Rev. B},
  volume = {87},
  issue = {10},
  pages = {104402},
  numpages = {5},
  year = {2013},
  doi = {10.1103/PhysRevB.87.104402},
  url = {https://link.aps.org/doi/10.1103/PhysRevB.87.104402}
}

@article{Matsuda_2010,
  title = {{Disordered Ground State and Magnetic Field-Induced Long-Range Order in an $S=3/2$ Antiferromagnetic Honeycomb Lattice Compound ${\mathrm{Bi}}_{3}{\mathrm{Mn}}_{4}{\mathrm{O}}_{12}({\mathrm{NO}}_{3})$}},
  author = {Matsuda, M. and Azuma, M. and Tokunaga, M. and Shimakawa, Y. and Kumada, N.},
  journal = {Phys. Rev. Lett.},
  volume = {105},
  issue = {18},
  pages = {187201},
  numpages = {4},
  year = {2010},
  doi = {10.1103/PhysRevLett.105.187201},
  url = {https://link.aps.org/doi/10.1103/PhysRevLett.105.187201}
}

@article{Lee_2020,
  title = {{Unconventional spin excitations in the $S=\frac{3}{2}$ triangular antiferromagnet ${\mathrm{RbAg}}_{2}\mathrm{Cr}{[{\mathrm{VO}}_{4}]}_{2}$}},
  author = {Lee, S. and Klauer, R. and Menten, J. and Lee, W. and Yoon, S. and Luetkens, H. and Lemmens, P. and M\"oller, A. and Choi, K.-Y.},
  journal = {Phys. Rev. B},
  volume = {101},
  issue = {22},
  pages = {224420},
  numpages = {7},
  year = {2020},
  doi = {10.1103/PhysRevB.101.224420},
  url = {https://link.aps.org/doi/10.1103/PhysRevB.101.224420}
}

@article{Collins_1997,
author = {Collins, M. F. and Petrenko, O. A.},
title = {Review/Synth\'ese: Triangular antiferromagnets},
journal = {Canadian Journal of Physics},
volume = {75},
number = {9},
pages = {605-655},
year = {1997},
doi = {10.1139/p97-007},
URL = {https://doi.org/10.1139/p97-007}
}

@article{Tapp_2017,
  title = {From magnetic order to spin-liquid ground states on the $S=\frac{3}{2}$ triangular lattice},
  author = {Tapp, J. and dela Cruz, C. R. and Bratsch, M. and Amuneke, N. E. and Postulka, L. and Wolf, B. and Lang, M. and Jeschke, H. O. and Valent\'{\i}, R. and Lemmens, P. and M\"oller, A.},
  journal = {Phys. Rev. B},
  volume = {96},
  issue = {6},
  pages = {064404},
  numpages = {7},
  year = {2017},
  doi = {10.1103/PhysRevB.96.064404},
  url = {https://link.aps.org/doi/10.1103/PhysRevB.96.064404}
}

@article{Zhong_2019,
author = {Ruidan Zhong  and Shu Guo  and Guangyong Xu  and Zhijun Xu  and Robert J. Cava },
title = {Strong quantum fluctuations in a quantum spin liquid candidate with a {Co}-based triangular lattice},
journal = {Proceedings of the National Academy of Sciences},
volume = {116},
number = {29},
pages = {14505-14510},
year = {2019},
doi = {10.1073/pnas.1906483116},
URL = {https://www.pnas.org/doi/abs/10.1073/pnas.1906483116}
}

@article{Li_2019,
  title = {Quantum spin state transitions in the spin-1 equilateral triangular lattice antiferromagnet {Na$_2$BaNi(PO$_4$)$_2$}},
  author = {Li, N. and Huang, Q. and Brassington, A. and Yue, X. Y. and Chu, W. J. and Guang, S. K. and Zhou, X. H. and Gao, P. and Feng, E. X. and Cao, H. B. and Choi, E. S. and Sun, Y. and Li, Q. J. and Zhao, X. and Zhou, H. D. and Sun, X. F.},
  journal = {Phys. Rev. B},
  volume = {104},
  issue = {10},
  pages = {104403},
  numpages = {9},
  year = {2021},
  doi = {10.1103/PhysRevB.104.104403},
  url = {https://link.aps.org/doi/10.1103/PhysRevB.104.104403}
}

@article{Xiang_2024,
  title = {Giant magnetocaloric effect in spin supersolid candidate {Na$_2$BaCo(PO$_4$)$_2$}},
  volume = {625},
  ISSN = {1476-4687},
  url = {http://dx.doi.org/10.1038/s41586-023-06885-w},
  doi = {10.1038/s41586-023-06885-w},
  number = {7994},
  journal = {Nature},
  publisher = {Springer Science and Business Media LLC},
  author = {Xiang,  Junsen and Zhang,  Chuandi and Gao,  Yuan and Schmidt,  Wolfgang and Schmalzl,  Karin and Wang,  Chin-Wei and Li,  Bo and Xi,  Ning and Liu,  Xin-Yang and Jin,  Hai and Li,  Gang and Shen,  Jun and Chen,  Ziyu and Qi,  Yang and Wan,  Yuan and Jin,  Wentao and Li,  Wei and Sun,  Peijie and Su,  Gang},
  year = {2024},
  pages = {270–275}
}

@article{Gao_2022,
  title = {Spin supersolidity in nearly ideal easy-axis triangular quantum antiferromagnet {Na$_2$BaCo(PO$_4$)$_2$}},
  volume = {7},
  pages = {89},
  ISSN = {2397-4648},
  number = {1},
  journal = {npj Quantum Materials},
  publisher = {Springer Science and Business Media LLC},
  author = {Gao,  Yuan and Fan,  Yu-Chen and Li,  Han and Yang,  Fan and Zeng,  Xu-Tao and Sheng,  Xian-Lei and Zhong,  Ruidan and Qi,  Yang and Wan,  Yuan and Li,  Wei},
  year = {2022},
  url = {http://dx.doi.org/10.1038/s41535-022-00500-3},
  doi = {10.1038/s41535-022-00500-3}
}

@article{Sheng_2022,
title = {Two-dimensional quantum universality in the spin-1/2 triangular-lattice quantum antiferromagnet {Na$_2$BaCo(PO$_4$)$_2$}},
author = {Jieming Sheng  and Le Wang  and Andrea Candini  and Wenrui Jiang  and Lianglong Huang  and Bin Xi  and Jize Zhao  and Han Ge  and Nan Zhao  and Ying Fu  and Jun Ren  and Jiong Yang  and Ping Miao  and Xin Tong  and Dapeng Yu  and Shanmin Wang  and Qihang Liu  and Maiko Kofu  and Richard Mole  and Giorgio Biasiol  and Dehong Yu  and Igor A. Zaliznyak  and Jia-Wei Mei  and Liusuo Wu },
journal = {Proceedings of the National Academy of Sciences},
volume = {119},
number = {51},
pages = {e2211193119},
year = {2022},
doi = {10.1073/pnas.2211193119},
URL = {https://www.pnas.org/doi/abs/10.1073/pnas.2211193119}
}

@article{Sheng_2025,
  title = {{Bose–Einstein condensation of a two-magnon bound state in a spin-1 triangular lattice}},
  ISSN = {1476-4660},
  url = {http://dx.doi.org/10.1038/s41563-024-02071-z},
  DOI = {10.1038/s41563-024-02071-z},
  journal = {Nature Materials},
  publisher = {Springer Science and Business Media LLC},
  author = {Sheng,  Jieming and Mei,  Jia-Wei and Wang,  Le and Xu,  Xiaoyu and Jiang,  Wenrui and Xu,  Lei and Ge,  Han and Zhao,  Nan and Li,  Tiantian and Candini,  Andrea and Xi,  Bin and Zhao,  Jize and Fu,  Ying and Yang,  Jiong and Zhang,  Yuanzhu and Biasiol,  Giorgio and Wang,  Shanmin and Zhu,  Jinlong and Miao,  Ping and Tong,  Xin and Yu,  Dapeng and Mole,  Richard and Cui,  Yi and Ma,  Long and Zhang,  Zhitao and Ouyang,  Zhongwen and Tong,  Wei and Podlesnyak,  Andrey and Wang,  Ling and Ye,  Feng and Yu,  Dehong and Yu,  Weiqiang and Wu,  Liusuo and Wang,  Zhentao},
  volume = {24},
  pages = {544},
  year = {2025}
}

@article{Zhang_2024,
  title = {Successive magnetic transitions in the $\text{spin-}\frac{5}{2}$ easy-axis triangular-lattice antiferromagnet {Na$_2$BaMn(PO$_4$)$_2$}: A neutron diffraction study},
  author = {Zhang, Chuandi and Xiang, Junsen and Su, Cheng and Sheptyakov, Denis and Liu, Xinyang and Gao, Yuan and Sun, Peijie and Li, Wei and Su, Gang and Jin, Wentao},
  journal = {Phys. Rev. B},
  volume = {110},
  issue = {21},
  pages = {214405},
  numpages = {8},
  year = {2024},
  doi = {10.1103/PhysRevB.110.214405},
  url = {https://link.aps.org/doi/10.1103/PhysRevB.110.214405}
}

@article{Nenert2020,
url = {https://doi.org/10.1515/zkri-2020-0028},
title = {{Crystal structure of the synthetic analogue of iwateite, Na$_2$BaMn(PO$_4$)$_2$: an X-ray powder diffraction and Raman study}},
author = {Gwilherm N\'enert and M. Mangir Murshed and Teycir Ben Hamed and Thorsten M. Gesing and Mongi Ben Amara},
pages = {433--437},
volume = {235},
number = {10},
journal = {Zeitschrift f\"ur Kristallographie - Crystalline Materials},
doi = {doi:10.1515/zkri-2020-0028},
year = {2020}
}

@article{Kim_2022,
doi = {10.1088/1361-648X/ac965f},
url = {https://dx.doi.org/10.1088/1361-648X/ac965f},
year = {2022},
publisher = {IOP Publishing},
volume = {34},
number = {47},
pages = {475803},
author = {Kim, Jaewook and Kim, Kyoo and Choi, Eunsang and Joon Ko, Young and Woo Lee, Dong and Ho Lim, Sang and Hoon Jung, Jong and Lee, Seungsu},
title = {{Magnetic phase diagram of a 2-dimensional triangular lattice antiferromagnet Na$_2$BaMn(PO$_4$)$_2$}},
journal = {Journal of Physics: Condensed Matter}
}

@article{Daisuke_2014,
  title={{Iwateite, Na$_2$BaMn(PO$_4$)$_2$, a new mineral from the Tanohata mine, Iwate Prefecture, Japan}},
  author={Daisuke Nishio-Hamane and Tetsuo Minakawa and Hanako Okada},
  journal={Journal of Mineralogical and Petrological Sciences},
  volume={109},
  number={1},
  pages={34-37},
  year={2014},
  doi={10.2465/jmps.131020a}
}

@article{Zhang_2024b,
  title = {Structural, magnetic, and magnetocaloric properties of triangular-lattice transition-metal phosphates},
  author = {Zhang, Chuandi and Xiang, Junsen and Zhu, Quanliang and Wu, Longfei and Zhang, Shanfeng and Xu, Juping and Yin, Wen and Sun, Peijie and Li, Wei and Su, Gang and Jin, Wentao},
  journal = {Phys. Rev. Mater.},
  volume = {8},
  issue = {4},
  pages = {044409},
  numpages = {9},
  year = {2024},
  doi = {10.1103/PhysRevMaterials.8.044409},
  url = {https://link.aps.org/doi/10.1103/PhysRevMaterials.8.044409}
}

@article{Kajita_2024,
author = {Kajita, Yoichi and Nagai, Takayuki and Yamagishi, Shigetada and Kimura, Kenta and Hagihala, Masato and Kimura, Tsuyoshi},
title = {{Ferroaxial Transitions in Glaserite-Type Na$_2$BaM(PO$_4$)$_2$ (M = Mg, Mn, Co, and Ni)}},
journal = {Chemistry of Materials},
volume = {36},
number = {15},
pages = {7451-7458},
year = {2024},
doi = {10.1021/acs.chemmater.4c01406},
URL = {https://doi.org/10.1021/acs.chemmater.4c01406}
}

@article{Li_2020,
  title = {Possible itinerant excitations and quantum spin state transitions in the effective spin-1/2 triangular-lattice antiferromagnet {Na$_2$BaCo(PO$_4$)$_2$}},
  volume = {11},
  pages = {4216 },
  ISSN = {2041-1723},
  url = {http://dx.doi.org/10.1038/s41467-020-18041-3},
  DOI = {10.1038/s41467-020-18041-3},
  number = {1},
  journal = {Nature Communications},
  publisher = {Springer Science and Business Media LLC},
  author = {Li,  N. and Huang,  Q. and Yue,  X. Y. and Chu,  W. J. and Chen,  Q. and Choi,  E. S. and Zhao,  X. and Zhou,  H. D. and Sun,  X. F.},
  year = {2020}
}

@article{Ishii_2011,
doi = {10.1209/0295-5075/94/17001},
url = {https://dx.doi.org/10.1209/0295-5075/94/17001},
year = {2011},
volume = {94},
number = {1},
pages = {17001},
author = {Ishii, R. and Tanaka, S. and Onuma, K. and Nambu, Y. and Tokunaga, M. and Sakakibara, T. and Kawashima, N. and Maeno, Y. and Broholm, C. and P. Gautreaux, D. and Chan, J. Y. and Nakatsuji, S.},
title = {Successive phase transitions and phase diagrams for the quasi-two-dimensional easy-axis triangular antiferromagnet {Rb$_4$Mn(MoO$_4$)$_3$}},
journal = {Europhysics Letters}
}

@article{Svistov_2006,
  title = {Magnetic phase diagram, critical behavior, and two-dimensional to three-dimensional crossover in the triangular lattice antiferromagnet {RbFe(MoO$_4$)$_2$}},
  author = {Svistov, L. E. and Smirnov, A. I. and Prozorova, L. A. and Petrenko, O. A. and Micheler, A. and B\"uttgen, N. and Shapiro, A. Ya. and Demianets, L. N.},
  journal = {Phys. Rev. B},
  volume = {74},
  issue = {2},
  pages = {024412},
  numpages = {10},
  year = {2006},
  doi = {10.1103/PhysRevB.74.024412},
  url = {https://link.aps.org/doi/10.1103/PhysRevB.74.024412}
}

@article{Seabra_2011,
  title = {Phase diagram of the classical {Heisenberg} antiferromagnet on a triangular lattice in an applied magnetic field},
  author = {Seabra, Luis and Momoi, Tsutomu and Sindzingre, Philippe and Shannon, Nic},
  journal = {Phys. Rev. B},
  volume = {84},
  issue = {21},
  pages = {214418},
  numpages = {14},
  year = {2011},
  doi = {10.1103/PhysRevB.84.214418},
  url = {https://link.aps.org/doi/10.1103/PhysRevB.84.214418}
}

@article{Yamamoto_2014,
  title = {Quantum Phase Diagram of the Triangular-Lattice {$XXZ$} Model in a Magnetic Field},
  author = {Yamamoto, Daisuke and Marmorini, Giacomo and Danshita, Ippei},
  journal = {Phys. Rev. Lett.},
  volume = {112},
  issue = {12},
  pages = {127203},
  numpages = {5},
  year = {2014},
  doi = {10.1103/PhysRevLett.112.127203},
  url = {https://link.aps.org/doi/10.1103/PhysRevLett.112.127203}
}

@article{Villain_1980,
	author = {Villain, J. and Bidaux, R. and Carton, J.-P. and Conte, R.},
	title = {Order as an effect of disorder},
	DOI= "10.1051/jphys:0198000410110126300",
	url= "https://doi.org/10.1051/jphys:0198000410110126300",
	journal = {J. Phys. France},
	year = {1980},
	volume = {41},
	number = {11},
	pages = {1263-1272}
}

@article{Henley1989,
  title = {Ordering due to disorder in a frustrated vector antiferromagnet},
  author = {Henley, Christopher L.},
  journal = {Phys. Rev. Lett.},
  volume = {62},
  issue = {17},
  pages = {2056--2059},
  numpages = {0},
  year = {1989},
  doi = {10.1103/PhysRevLett.62.2056},
  url = {https://link.aps.org/doi/10.1103/PhysRevLett.62.2056}
}

@article{Qureshi2019,
author = "Qureshi, Navid",
title = "{{\it Mag2Pol}: a program for the analysis of spherical neutron polarimetry, flipping ratio and integrated intensity data}",
journal = "Journal of Applied Crystallography",
year = "2019",
volume = "52",
number = "1",
pages = "175--185",
doi = {10.1107/S1600576718016084},
url = {https://doi.org/10.1107/S1600576718016084}
}

@article{Scheie_2018,
title={{LongHCPulse: Long-pulse heat capacity on a Quantum Design PPMS}},
author={Scheie, Allen},
journal={Journal of Low Temperature Physics},
volume={193},
pages={60--73},
year={2018},
doi={10.1007/s10909-018-2042-9},
url = {https://doi.org/10.1007/s10909-018-2042-9}
}

@article{data_D23,
  year     = {2024},
  url      = {https://doi.ill.fr/10.5291/ILL-DATA.CRG-3064},
  journal  = {https://doi.ill.fr/10.5291/ILL-DATA.CRG-3064}
}

@article{data_D23b,
  year     = {2024},
  url      = {https://doi.ill.fr/10.5291/ILL-DATA.5-41-1252},
  journal  = {https://doi.ill.fr/10.5291/ILL-DATA.5-41-1252}
}

@article{dos_santos_spin-resolved_2018,
	title = {Spin-resolved inelastic electron scattering by spin waves in noncollinear magnets},
	volume = {97},
	url = {https://link.aps.org/doi/10.1103/PhysRevB.97.024431},
	doi = {10.1103/PhysRevB.97.024431},
	number = {2},
	urldate = {2018-01-26},
	journal = {Physical Review B},
	author = {dos Santos, Flaviano José and dos Santos Dias, Manuel and Guimarães, Filipe Souza Mendes and Bouaziz, Juba and Lounis, Samir},
	year = {2018},
	pages = {024431}
}

@article{Kimura_2008,
  title = {Magnetoelectric control of spin-chiral ferroelectric domains in a triangular lattice antiferromagnet},
  author = {Kimura, Kenta and Nakamura, Hiroyuki and Ohgushi, Kenya and Kimura, Tsuyoshi},
  journal = {Phys. Rev. B},
  volume = {78},
  issue = {14},
  pages = {140401},
  numpages = {4},
  year = {2008},
  doi = {10.1103/PhysRevB.78.140401},
  url = {https://link.aps.org/doi/10.1103/PhysRevB.78.140401}
}

@article{Kitazawa_1999,
title = {High-field magnetization of triangular lattice antiferromagnet: {GdPd$_2$Al$_3$}},
journal = {Physica B: Condensed Matter},
volume = {259-261},
pages = {890-891},
year = {1999},
issn = {0921-4526},
doi = {https://doi.org/10.1016/S0921-4526(98)01101-6},
url = {https://www.sciencedirect.com/science/article/pii/S0921452698011016},
author = {H Kitazawa and H Suzuki and H Abe and J Tang and G Kido}
}

@misc{Supplemental,
      note = {See Supplemental Material for supporting experimental and theoretical information.}
}

@article{Hearmon_2012,
  title = {{Electric Field Control of the Magnetic Chiralities in Ferroaxial Multiferroic RbFe(MoO$_4$)$_2$}},
  author = {Hearmon, Alexander J. and Fabrizi, Federica and Chapon, Laurent C. and Johnson, R. D. and Prabhakaran, Dharmalingam and Streltsov, Sergey V. and Brown, P. J. and Radaelli, Paolo G.},
  journal = {Phys. Rev. Lett.},
  volume = {108},
  issue = {23},
  pages = {237201},
  numpages = {5},
  year = {2012},
  doi = {10.1103/PhysRevLett.108.237201},
  url = {https://link.aps.org/doi/10.1103/PhysRevLett.108.237201}
}

@article{Jia_2024,
  title = {{Quantum spin supersolid as a precursory Dirac spin liquid in a triangular lattice antiferromagnet}},
  author = {Jia, Haichen and Ma, Bowen and Wang, Z. D. and Chen, Gang},
  journal = {Phys. Rev. Res.},
  volume = {6},
  issue = {3},
  pages = {033031},
  numpages = {12},
  year = {2024},
  doi = {10.1103/PhysRevResearch.6.033031},
  url = {https://link.aps.org/doi/10.1103/PhysRevResearch.6.033031}
}

@article{Sakhratov_2022,
  title = {{High-field magnetic structure of the triangular antiferromagnet RbFe(MoO$_{4}$)$_{2}$}},
  author = {Sakhratov, Yu. A. and Prokhnenko, O. and Shapiro, A. Ya. and Zhou, H. D. and Svistov, L. E. and Reyes, A. P. and Petrenko, O. A.},
  journal = {Phys. Rev. B},
  volume = {105},
  issue = {1},
  pages = {014431},
  numpages = {12},
  year = {2022},
  doi = {10.1103/PhysRevB.105.014431},
  url = {https://link.aps.org/doi/10.1103/PhysRevB.105.014431}
}

@article{White_2013,
  title = {{Multiferroicity in the generic easy-plane triangular lattice antiferromagnet RbFe(MoO$_{4}$)$_{2}$}},
  author = {White, J. S. and Niedermayer, Ch. and Gasparovic, G. and Broholm, C. and Park, J. M. S. and Shapiro, A. Ya. and Demianets, L. A. and Kenzelmann, M.},
  journal = {Phys. Rev. B},
  volume = {88},
  issue = {6},
  pages = {060409},
  numpages = {5},
  year = {2013},
  doi = {10.1103/PhysRevB.88.060409},
  url = {https://link.aps.org/doi/10.1103/PhysRevB.88.060409}
}

@misc{huang_2025u,
      title={Universal dynamics of a pair condensate}, 
      author={Qing Huang and Hao Zhang and Yiqing Hao and Weiliang Yao and Daniel M. Pajerowski and Adam A. Aczel and Eun Sang Choi and Kipton Barros and Bruce Normand and Haidong Zhou and Andreas M. L\"auchli and Xiaojian Bai and Shang-Shun Zhang},
      year={2025},
      eprint={2503.13609},
      archivePrefix={arXiv},
      url={https://arxiv.org/abs/2503.13609}
}

@misc{data, 
      title={Reproducibility package for article: Spin structures and phase diagrams of the spin-5/2 triangular-lattice antiferromagnet  {{Na}$_2${BaMn}({PO}$_4$)$_2$} under magnetic field}, 
      url={https://doi.org/10.6084/m9.figshare.29899553}, 
      doi={10.6084/m9.figshare.29899553}, 
      publisher={figshare}, 
      howpublished = {{Figshare}},
      author={Biniskos, Nikolaos and dos Santos, Flaviano José and Stekiel, Michal and Schmalzl, Karin and Ressouche, Eric and Sviták, David and Labh, Ankit and Vališka, Michal and Marzari, Nicola and Čermák, Petr}, 
      year={2025}, 
      month={Aug} 
}

@article{Starykh_2015,
  author  = {Oleg A. Starykh},
  title   = {Unusual ordered phases of highly frustrated magnets},
  journal = {Reports on Progress in Physics},
  year    = {2015},
  volume  = {78},
  number  = {5},
  pages   = {052502},
  doi     = {10.1088/0034-4885/78/5/052502},
  url = {https://doi.org/10.1088/0034-4885/78/5/052502}
}

@article{Kawamura_2010,
  author  = {Hikaru Kawamura and Atsushi Yamamoto and Tsuyoshi Okubo},
  title   = {{$Z_2$-Vortex Ordering of the Triangular-Lattice Heisenberg Antiferromagnet}},
  journal = {Journal of the Physical Society of Japan},
  year    = {2010},
  volume  = {79},
  number  = {2},
  pages   = {023701},
  doi     = {10.1143/JPSJ.79.023701},
  url = {https://doi.org/10.1143/JPSJ.79.023701}
}

@article{Mun_2014,
  author  = {Mun, Eundeok and Frontzek, M. and Podlesnyak, A. and Ehlers, G. and Barilo, S. and Shiryaev, S. V. and Zapf, Vivien S.},
  title   = {{High magnetic field evolution of ferroelectricity in CuCrO$_2$}},
  journal = {Phys. Rev. B},
  volume = {89},
  issue = {5},
  pages = {054411},
  numpages = {9},
  year = {2014},
  doi = {10.1103/PhysRevB.89.054411},
  url = {https://link.aps.org/doi/10.1103/PhysRevB.89.054411}
}

@article{Sakhratov_2016,
  author  = {Sakhratov, Yu. A. and Svistov, L. E. and Kuhns, P. L. and Zhou, H. D. and Reyes, A. P.},
  title   = {{Magnetic phases of the quasi-two-dimensional antiferromagnet CuCrO$_2$ on a triangular lattice}},
  journal = {Phys. Rev. B},
  volume = {94},
  issue = {9},
  pages = {094410},
  numpages = {8},
  year = {2016},
  doi = {10.1103/PhysRevB.94.094410},
  url = {https://link.aps.org/doi/10.1103/PhysRevB.94.094410}
}

@article{Frontzek_2011,
  author  = {Frontzek, M. and Haraldsen, J. T. and Podlesnyak, A. and Matsuda, M. and Christianson, A. D. and Fishman, R. S. and Sefat, A. S. and Qiu, Y. and Copley, J. R. D. and Barilo, S. and Shiryaev, S. V. and Ehlers, G.},
  title   = {{Magnetic excitations in the geometrically frustrated multiferroic CuCrO$_2$}},
  journal = {Phys. Rev. B},
  volume = {84},
  issue = {9},
  pages = {094448},
  numpages = {7},
  year = {2011},
  doi = {10.1103/PhysRevB.84.094448},
  url = {https://link.aps.org/doi/10.1103/PhysRevB.84.094448}
}

\end{document}